\documentclass[draft]{agujournal2019}
\usepackage{url} 
\usepackage{amsmath}
\usepackage{soul}
\draftfalse
\journalname{Journal of Advances in Modeling Earth Systems (JAMES)}

\begin{document}

\title{Deep learning of systematic sea ice model errors from data assimilation increments}

\authors{William Gregory\affil{1}, Mitchell Bushuk\affil{2}, Alistair Adcroft\affil{1}, Yongfei Zhang\affil{1}, Laure Zanna\affil{3}}

\affiliation{1}{Atmospheric and Oceanic Sciences Program, Princeton University, NJ, USA}
\affiliation{2}{Geophysical Fluid Dynamics Laboratory, NOAA, Princeton, NJ, USA}
\affiliation{3}{Courant Institute of Mathematical Sciences, New York University, New York, NY, USA}

\correspondingauthor{Will Gregory}{wg4031@princeton.edu}

\begin{keypoints} 
\item We show that sea ice data assimilation increments closely reflect the systematic bias patterns of a global ice-ocean model
\item Convolutional neural networks can make skillful predictions of sea ice data assimilation increments, using only model state variables
\item The skillful predictions suggest the network could be used as a parameterization to reduce sea ice biases in free-running model simulations
\end{keypoints}

\begin{abstract} 
Data assimilation is often viewed as a framework for correcting short-term error growth in dynamical climate model forecasts. When viewed on the time scales of climate however, these short-term corrections, or analysis increments, can closely mirror the systematic bias patterns of the dynamical model. In this study, we use convolutional neural networks (CNNs) to learn a mapping from model state variables to analysis increments, in order to showcase the feasibility of a data-driven model parameterization which can predict state-dependent model errors. We undertake this problem using an ice-ocean data assimilation system within the Seamless system for Prediction and EArth system Research (SPEAR) model, developed at the Geophysical Fluid Dynamics Laboratory, which assimilates satellite observations of sea ice concentration every 5 days between 1982--2017. The CNN then takes inputs of data assimilation forecast states and tendencies, and makes predictions of the corresponding sea ice concentration increments. Specifically, the inputs are states and tendencies of sea ice concentration, sea-surface temperature, ice velocities, ice thickness, net shortwave radiation, ice-surface skin temperature, sea-surface salinity, as well as a land-sea mask. We find the CNN is able to make skillful predictions of the increments in both the Arctic and Antarctic and across all seasons, with skill that consistently exceeds that of a climatological increment prediction. This suggests that the CNN could be used to reduce sea ice biases in free-running SPEAR simulations, either as a sea ice parameterization or an online bias correction tool for numerical sea ice forecasts.

\end{abstract}

\section*{Plain Language Summary} 
To make predictions of the Earth's climate system we use expensive computer simulations, called climate models. These models are not perfect however, as we often need to approximate certain physical laws in order to save on compute time. On the other hand we have observational climate data, however these data have limited space and time coverage and also contain errors because of noise and assumptions about how our measurements relate to the quantity we are interested in. Therefore we often use a process called data assimilation to combine our climate model predictions together with observations, to produce our `best guess' of the climate system. The difference between our best-guess-model and our original climate model prediction then gives us clues as to how wrong our original climate model is. In this work we use some fancy statistics, called machine learning, where we show a computer algorithm lots of examples of sea ice, atmosphere and ocean climate model predictions, and see if it can learn its own inherent sea ice errors. We find that it can do this well, which means that we can hopefully incorporate the machine learning algorithm into the original climate model to improve its future climate predictions.

\section{Introduction}\label{sect:intro}
The influence of structural errors within climate models due to missing physics, imperfect parameterizations of subgrid-scale processes, as well as errors in the underlying numerics, leads to systematic biases across the atmosphere, land, sea ice, and ocean. Subsequently, our ability to diagnose and correct these biases ultimately governs the accuracy of numerical weather and climate predictions on different time scales \cite{Stevens2013}. In the context of sea ice for example, much effort has been afforded to the improvement of model physics and subgrid parameterizations through the development of e.g., ice thickness distribution \cite{Thorndike1975,Bitz2001} and floe-size distribution theory \cite{Rothrock1984,Horvat2015}, surface melt-pond \cite{Flocco2012}, ice drift \cite{Tsamados2013} and lateral melt parameterizations \cite{Smith2022}, as well as sea ice rheology \cite{Hibler1979,Dansereau2016,Olason2022}. Such studies have shown how the improved representation of sea ice physics produces model simulations which more closely reflect observations in terms of either their mean sea ice volume, drift, or ice thickness distribution. Despite this, however, biases will often persist due to the fact that physical processes must be approximated in order to meet computational restraints, and that parameterizations are often based on sparse observations which were collected under a climate regime which may not generalize to future conditions \cite{Notz2012}. Sea ice is also strongly coupled to both the atmosphere and ocean via mechanical and thermodynamic forcing, thus sea ice biases can also manifest from biases in these components.

Many previous studies have leveraged data assimilation (DA) as a way to either assess model error or better understand model physics within numerical weather prediction (NWP) systems \cite{Leith1978,Klinker1992,Dee2005,Rodwell2007,Palmer2011,Carrassi2011,Mitchell2015,Crawford2020,Laloyaux2020}. Generally, DA can be considered a Bayesian framework for combining a model forecast with observations in order to produce an optimal estimate of a given set of climate state variables, often called the \emph{analysis state}. The difference between this analysis state and the model forecast prior to assimilation is then the \emph{analysis increment}, which represents our `best guess' as to the appropriate correction to the model forecast when taking into account both model and observational uncertainty. One caveat to this is that many DA systems do not formally account for systematic model biases, and so these systems often produce non-zero values in the time-mean of their analysis increments; indicating consistent discrepancies between the model and observations. Attributing such errors to their correct source is also non-trivial \cite{Dee2004,Dee2005}, as model biases can manifest non-locally in space and time \cite{Palmer2011,Wang2014} and involve non-linear interactions across different model components \cite{Large2006,Kim2022}. Observations themselves may also contain systematic errors, such as the design of weather filters in satellite-derived sea ice area retrievals \cite{Kern2019} and uncertainties related to summer ice surface properties \cite{Kern2020}. While some studies have shown relative success in separating systematic errors between observations and models \cite{Auligne2007,Dee2009}, many assimilation systems simply assume that the observational errors are uncorrelated and Gaussian, and subsequently any systematic patterns within the analysis increments can largely be considered a manifestation of the various model biases. Under this assumption, the increments can be seen as a reflection of model error growth associated with missing or imbalanced physical processes occurring over short time scales, often called \emph{fast physics} errors, however such errors ultimately have an impact on the model's bias patterns over climate time scales as well \cite{Murphy2004,Rodwell2007}.

The analysis increments therefore provide useful information on model deficiencies, which could inform new parameterizations to reduce systematic model biases. Indeed, variational schemes such as weak-constraint 4D-Var \cite{Wergen1992,Zupanski1993,Tremolet2007,Laloyaux2020} already aim to account for systematic model error during DA, and while this is invaluable in NWP, the underlying model physics remains unchanged, meaning that a free-running model simulation invariably remains biased. An alternative approach which has been explored in the ocean modeling community \cite{Chepurin2005,Balmaseda2007,Lu2020} is to use DA to first derive the climatological components of the systematic model biases, and then incorporate these components back into the model as an adjustment to the model state tendencies. \citeA{Lu2020} for example designed an ocean DA system which assimilates temperature and salinity profile data into version 6 of the Modular Ocean Model (MOM6), and from this they derived analysis increments of temperature and salinity at each model grid cell location and vertical level. They subsequently computed the daily climatology of the increments, which represent the systematic component of model error for each field on any given day of the year, and incorporated these as a three-dimensional adjustment to the model temperature and salinity tendencies for subsequent MOM6 ocean simulations. This `ocean tendency adjustment' was found to reduce ocean model bias and improve the skill of coupled model seasonal predictions of the El Ni\~no Southern Oscillation.

More recently, machine learning (ML) has been put forward as a data-driven framework for targeting model biases. ML, in particular deep learning (DL), algorithms have become increasingly popular in climate research for a variety of applications ranging from NWP \cite{Pathak2022,Bi2022} to satellite altimetry data processing \cite{Dawson2022,Landy2022}. In the context of dynamical climate models, DL algorithms have proven effective tools for deriving model parameterizations directly from numerical simulations. For example, many past studies have focused on learning subgrid parameterizations from high resolution experiments and/or observations of the ocean \cite{Bolton2019,Zanna2020,Zhu2022}, atmosphere \cite{Brenowitz2018,Gentine2018,Rasp2018,OGorman2018,Yuval2020,Wang2022}, and sea ice \cite{Finn2023}. In the context of DA-based approaches, some recent studies have relied on iterative sequences of DA and ML to infer unresolved scale parameterizations from sparse and noisy observations \cite{Brajard2021}, or to learn state-dependent model error from analysis increments \cite{Farchi2021} and nudging tendencies \cite{Watt2021,Bretherton2022}, while others have combined DA with equation discovery to extract interpretable structural model errors \cite{Mojgani2022}. Many of these studies have relied on idealized models to showcase the feasibility of various DA-ML methodologies, however recently \citeA{Bonavita2020} used ML to learn state-dependent model errors from atmospheric analysis increments produced from a 4D-Var simulation within the the Integrated Forecasting System (IFS) model at the European Centre for Medium-range Weather Forecasts (ECMWF), and similarly, \citeA{Laloyaux2022} attempted to learn atmospheric temperature errors within the same IFS model using the model bias directly, as a way to a-priori define the bias model within subsequent 4D-Var simulations. This latter approach however was unable to outperform the current operational weak-constraint 4D-Var system at ECMWF.

In this study, we present a DA-based ML approach to learn the systematic biases of a large-scale sea ice model used for climate simulations. We learn state-dependent sea ice errors within the  Seamless system for Prediction and EArth system Research (SPEAR) model \cite{Delworth2020}, developed at the Geophysical Fluid Dynamics Laboratory (GFDL), by constructing convolutional neural networks (CNNs) which learn a functional mapping from model state variables to sea ice DA increments. Somewhat different to previous studies which have been centered around DA and ML in idealized model contexts \cite{Brajard2021,Farchi2021,Mojgani2022}, our application here is, to our knowledge, the first example of using ML to learn systematic model error from DA increments in a global ice-ocean model (though similar approaches have previously been explored within large-scale atmospheric models \cite{Bonavita2020,Chen2022}). We also choose to learn sea ice errors from DA increments as opposed to learning the model bias directly (e.g., \citeA{Laloyaux2022}), as the increments have inherently accounted for model and observational uncertainty, and they also provide a full spatio-temporal record of errors for model state variables which are not direct observables, such as subgrid ice thickness distribution category concentrations. It is also worth noting that while we present this article in the context of using ML to make offline predictions of sea ice DA increments, we are ultimately working towards an ML model which can be implemented as an online sea ice parameterization within SPEAR. Similar to previous works \cite{Grundner2022,Wang2022}, this article is therefore an initial evaluation into the feasibility of this task, based on offline performance.

This paper is structured as follows: Section \ref{sect:ModConf} provides a brief overview of the SPEAR ice-ocean model configuration, as well as the sea ice DA setup. Section \ref{sect:incsvbias} then highlights how the climatological sea ice concentration (SIC) bias of a SPEAR ice-ocean model experiment maps closely onto the SPEAR SIC DA increments, motivating the idea of learning systematic model error from analysis increments. Section \ref{sect:CNNs} describes the ML problem setup and documents the CNN architectures and hyperparameter settings. Section \ref{sect:Results} then showcases the predictive performance of the CNN, and provides an assessment of the CNN sensitivity and generalization ability. Section \ref{sect:Disc} presents a discussion on the results and outlines considerations for future work relating to sea ice parameterizations and climate prediction. A final summary is then given in section \ref{sect:Conc}, as well as an outlook on the broader implications of this work within the climate modeling community.

\section{Model configuration}\label{sect:ModConf}
\subsection{SPEAR ice-ocean model}\label{sect:SPEAR}
SPEAR is a fully coupled ice-ocean-atmosphere-land model, with nominal 1$^\circ$ horizontal resolution in the ice and ocean components \cite{Delworth2020}. The SPEAR ocean component is based on MOM6, with 75 vertical layers, and the sea ice component on version 2 of the Sea Ice Simulator (SIS2; see \citeA{Adcroft2019} for details on both MOM6 and SIS2).  In this work, we consider an ice-ocean model configuration of SPEAR forced by atmospheric conditions and river runoff from the Japanese 55-year Reanalysis for driving ocean-sea-ice models (JRA55-do; \citeA{Tsujino2018}).

The SIS2 ice dynamics are solved using a elastic-viscous-plastic rheology on a tripolar Arakawa C-grid \cite{Bouillon2009}, with advection performed with a modified upwind scheme \cite{Adcroft2019}. The energy-conserving thermodynamics of the ice follows that of \citeA{Bitz1999}, and uses a vertical structure consisting of four ice layers and a single snow layer. Following \citeA{Bitz2001}, five ice thickness distribution categories are implemented in a Lagrangian scheme, with thickness boundaries of 0.1, 0.3, 0.7, 1.1 metres. The coupling between ice and ocean occurs at a frequency of 60 minutes, with a temperature coupling coefficient of 240 Wm$^{-2}$K$^{-1}$, while faster coupling with the atmosphere occurs through a surface skin temperature every 20 minutes. The model does not contain melt-pond, subgrid ridging, lateral melt, or land-fast ice parameterizations.

\subsection{Sea ice data assimilation and model experiments}\label{sect:DA}
An experimental ice-ocean DA system within SPEAR was recently developed by \citeA{Zhang2021}, whereby satellite-derived SIC from the National Snow and Ice Data Center (NSIDC; \citeA{Cavalieri1996}) NASA Team algorithm is assimilated into SIS2 via the Ensemble Adjustment Kalman Filter (EAKF; \citeA{Anderson2001}), and MOM6 sea-surface temperatures are nudged towards observations from version 2 of the 1$^\circ$ gridded Optimum Interpolation Sea-Surface Temperature (OISSTv2) data set \cite{Reynolds2007,Banzon2016}. In this section we give a brief overview of this DA setup, although the reader is referred to \citeA{Zhang2021} for further details.

A single SPEAR ensemble member is initialised in 1958 with World Ocean Atlas ocean conditions, and a prescribed atmosphere from JRA55-do reanalysis. This single member is then integrated forward to 1979 in order to `spin up' the ocean and sea ice, which then provides the initial ice and ocean conditions for a set of 30 ensemble members, each with individual perturbed sea ice physics. These perturbations correspond to independent random draws from a uniform distribution for sea ice model parameters including the ice strength parameter \cite{Hibler1979}, as well as the ice, snow, and pond albedo parameters \cite{Briegleb2007}. The distribution for ice strength spans 20,000--50,000 Nm$^{-1}$, while the distribution for albedo parameters spans -1.6--1.6 standard deviations \cite{Zhang2021}. The 30 perturbed physics ensemble members are then integrated forward from 1979 to 1982 in order to spin up the sea ice and generate sufficient spread across the ensemble. After which, the first sea ice DA update is made on January 6\textsuperscript{th} 1982 and continues every 5 days until December 27\textsuperscript{th} 2017, providing a total of 2618 assimilation cycles. This corresponds to 73 cycles per year except for 1982, 1987 and 1988, which contain 71, 68 and 70 cycles, respectively. There are 71 cycles in 1982 because the first cycle begins after the initial update on January 6\textsuperscript{th}, and 68 and 70 in 1987 and 1988 due to missing satellite observations between December 3\textsuperscript{rd} 1987 and January 13\textsuperscript{th} 1988 \cite{Cavalieri1996}. Note that, for convenience, the model is run with a `no leap' calendar which excludes leap-year days.

During each assimilation cycle, a model forecast is run until 00:00 hours UTC on the assimilation day (e.g., Jan 6\textsuperscript{th}), at which point the ice concentration from each of the model's five individual ice thickness distribution categories (hereafter SICN; note that SIC = $\sum_{k=1}^5 \mathrm{SICN}_k$) are passed to the EAKF, along with the satellite SIC observations, and subsequently the SICN forecasts are updated by the filter to produce their analysis states. Given that the aggregate SIC analysis state corresponds to the sum of the SICN analysis states, it is necessary to post-process SICN after each DA cycle in order avoid non-physical values in SIC, which is bounded between 0 and 1. This is achieved by appropriately scaling each of the SICN states when SIC is greater than 1, and setting SICN to 0 when SIC is negative. After post-processing, the analysis increments are then computed for each of the five category concentrations ($\Delta$SICN), and for each of the 30 ensemble members. State variables for each ensemble member are saved as daily mean fields during model integration, giving 365 days $\times$ 36 years $=$ 13140 daily forecasts for each variable. For the remainder of this article we consider only the ensemble mean fields for both the model state variables and the analysis increments. 

In order to understand the inherent SIC bias patterns within SPEAR, the next section includes a comparison of the SIC DA increments ($\Delta$SIC) to an additional model experiment without SIC DA, referred to here as FREE. This experiment corresponds to the same JRA-forced ice-ocean model configuration with sea-surface temperature nudging, as well as the same perturbed sea ice physics, and initial conditions from the spinup run as the SIC DA experiment. Therefore the FREE experiment configuration is identical to the SIC DA run, except for the assimilation of SIC observations.

\begin{figure}[t!]
    \centering
    \includegraphics[width=1\linewidth]{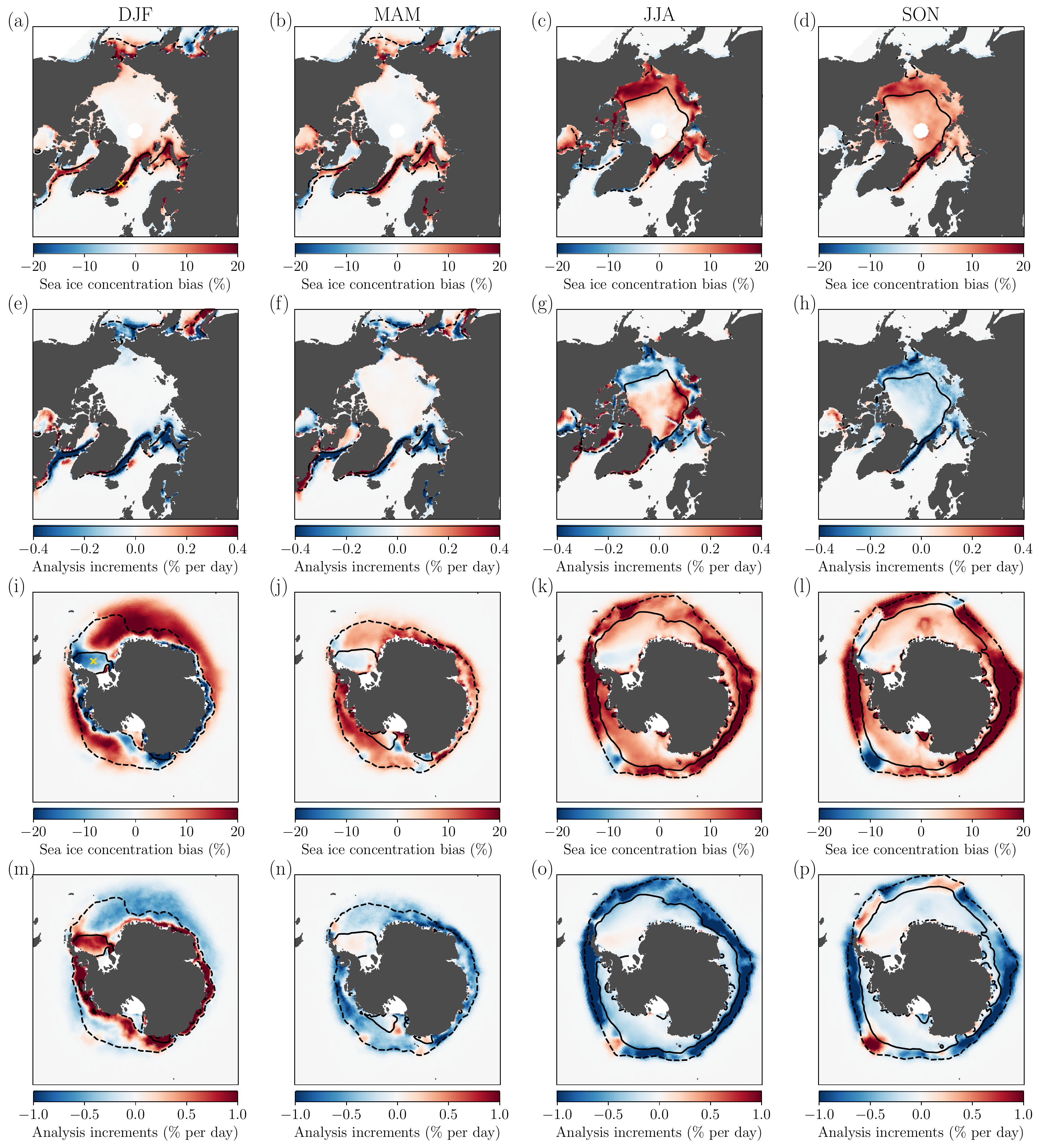}
    \caption{Seasonal climatologies of SPEAR free-running model bias (model minus observations) and sea ice concentration analysis increments, for both the Arctic (a)--(h) and Antarctic (i)--(p). Columns from left to right show DJF, MAM, JJA, SON climatologies, computed over the period 1982--2017. Dashed and solid contours denote the observed climatology marginal ice zone boundaries over the same period (15 and 75\% SIC contours, respectively). Yellow markers in (a) and (i) are example grid-point locations used for analysis in Figure \ref{fig:DAforecasts}.}
    \label{fig:incsvbias}
\end{figure}

\begin{figure}[t!]
    \centering
    \includegraphics[width=0.7\linewidth]{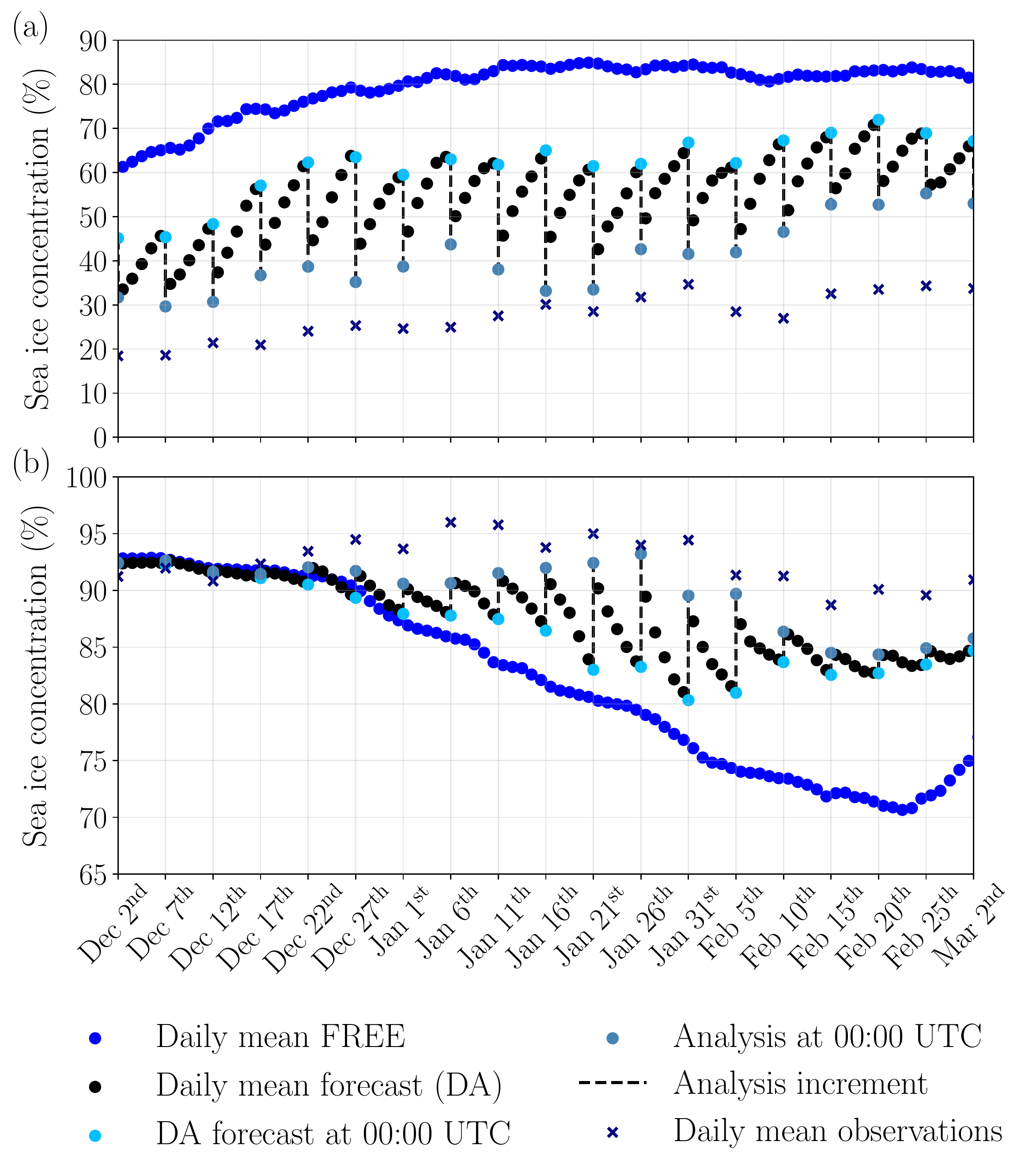}
    \caption{SPEAR sea ice concentration data assimilation example, shown for one grid cell as daily climatologies (1982--2017). Examples are presented for the Arctic (a) and Antarctic (b) through the period December--February. The grid cells for both the Arctic and Antarctic examples correspond to locations in the GIN Sea and Weddell Sea, respectively (see the yellow markers in Figures \ref{fig:incsvbias}a and \ref{fig:incsvbias}i).}
    \label{fig:DAforecasts}
\end{figure}

\section{Analysis increments and model bias in SPEAR}\label{sect:incsvbias}
Learning systematic model error from DA increments, with the goal of an eventual sea ice parameterization which reduces climate model bias, relies on the assumption that the fast physics errors captured within the DA increments reflect the long-term systematic biases of the free-running model \cite{Rodwell2007}. In this section, we examine whether this necessary condition is satisfied, making comparisons of $\Delta$SIC to the climatological bias of the FREE experiment. The model bias is computed relative to NSIDC NASA Team satellite SIC observations. 

Figure \ref{fig:incsvbias} shows seasonal climatologies of the SPEAR FREE SIC model bias and $\Delta$SIC between 1982--2017, for both the Arctic and Antarctic. Here we notice that the free-running model is, on average, positively biased in both hemispheres, with larger magnitude biases in the Antarctic. Crucially, we find largely consistent patterns between the model bias and $\Delta$SIC. In the Arctic for example, the large positive biases in the Greenland, Iceland, Norwegian (GIN) and Barents seas (east Atlantic) are mirrored by overall negative increments, hence the DA is acting to remove sea ice in this region. The winter Arctic SIC biases appear to be related to systematic biases in the sea ice edge position, which is apparent when noticing that the increments in the fully covered ice pack (north of the 75\% observed SIC contour) are relatively small compared to the marginal ice zones in DJF and MAM. The presence of larger increments in the central ice pack in JJA and SON are then likely a reflection of local SIC errors in the ice-covered zone in addition to ice edge position errors. The only notable discrepancy between model bias and $\Delta$SIC in the Arctic appears to be in the Kara and Laptev shelf seas in JJA, where both the model bias and increments are positive. This suggests that the assimilation forecasts are negatively biased in this region, which may be related to a residual overshooting problem in the DA experiment, as highlighted in the original SPEAR sea ice DA study by \citeA{Zhang2021}.

Turning to the Antarctic, despite largely positive biases across all seasons, negative biases dominate many of the coastal regions in the austral summer (DJF), including the Weddell Sea, whereby many of these biases become lower in magnitude or even positive by austral winter (JJA). Interestingly, the isolated negative bias towards the north-eastern edge of the Ross Sea is a persistent feature from MAM through to SON, reaching its largest magnitude in SON. This may be related to strong deep ocean convection in this region \cite{Adcroft2019}, which manifests as positively biased sea-surface temperatures which are co-located with the negatively biased SIC zone (see Figure \ref{fig:S1} in Supporting Information S1). Overall, the strong spatial and seasonal agreement between the free-running model bias and $\Delta$SIC supports this study's plan to use DA increments to learn a parameterization of sea ice model error.

Visualising the time evolution of the sea ice DA forecasts (Figure \ref{fig:DAforecasts}) shows the relationship between systematic biases and analysis increments more clearly. In the GIN Sea (Figure \ref{fig:DAforecasts}a), we can see that the model forecasts in each DA cycle (black dots) are drifting towards the positively-biased free-running model state (dark blue dots) over the 5-day forecast period, and as such the analysis increments (dashed black lines) are systematically negative to account for this. Similarly, in the Weddell Sea (Figure \ref{fig:DAforecasts}b) the forecasts are drifting towards the negatively-biased free-running model state, resulting in systematically positive increments. The forecast drift that is observed in either case can be quantified by the assimilation forecast tendencies, which for a given assimilation cycle $i$, corresponds to the time-derivative of the forecast $c$ at time $t$, or more simply $\dot{c}_i(t) = c_i(t) - c_i(t-1)$. The \emph{total} forecast tendency for a given assimilation cycle is then the sum of the individual daily tendencies: $\dot{c}_i(1) + \dot{c}_i(2) + ... + \dot{c}_i(5)$. \citeA{Klinker1992} showed that the mean total tendencies across a large number of assimilation cycles, referred to as the \emph{systematic forecast tendency}, is approximately equal to the negative of the analysis increments, which is also the case in our SPEAR DA experiments (see Figure \ref{fig:S2}). Building on this, \citeA{Rodwell2007} then later described how the forecast tendencies can be broken down into tendencies associated with the model's representation of various resolved and parameterized physical processes, and subsequently used them to make assessments of model physics errors after a model change had been made. In our study here, we utilize this inherent link between forecast tendencies and analysis increments to construct CNNs which use inputs of both state variables from the DA forecasts, as well as their associated forecast tendencies, in order to predict $\Delta$SICN.

\begin{figure}[t!]
    \centering
    \includegraphics[width=1\linewidth]{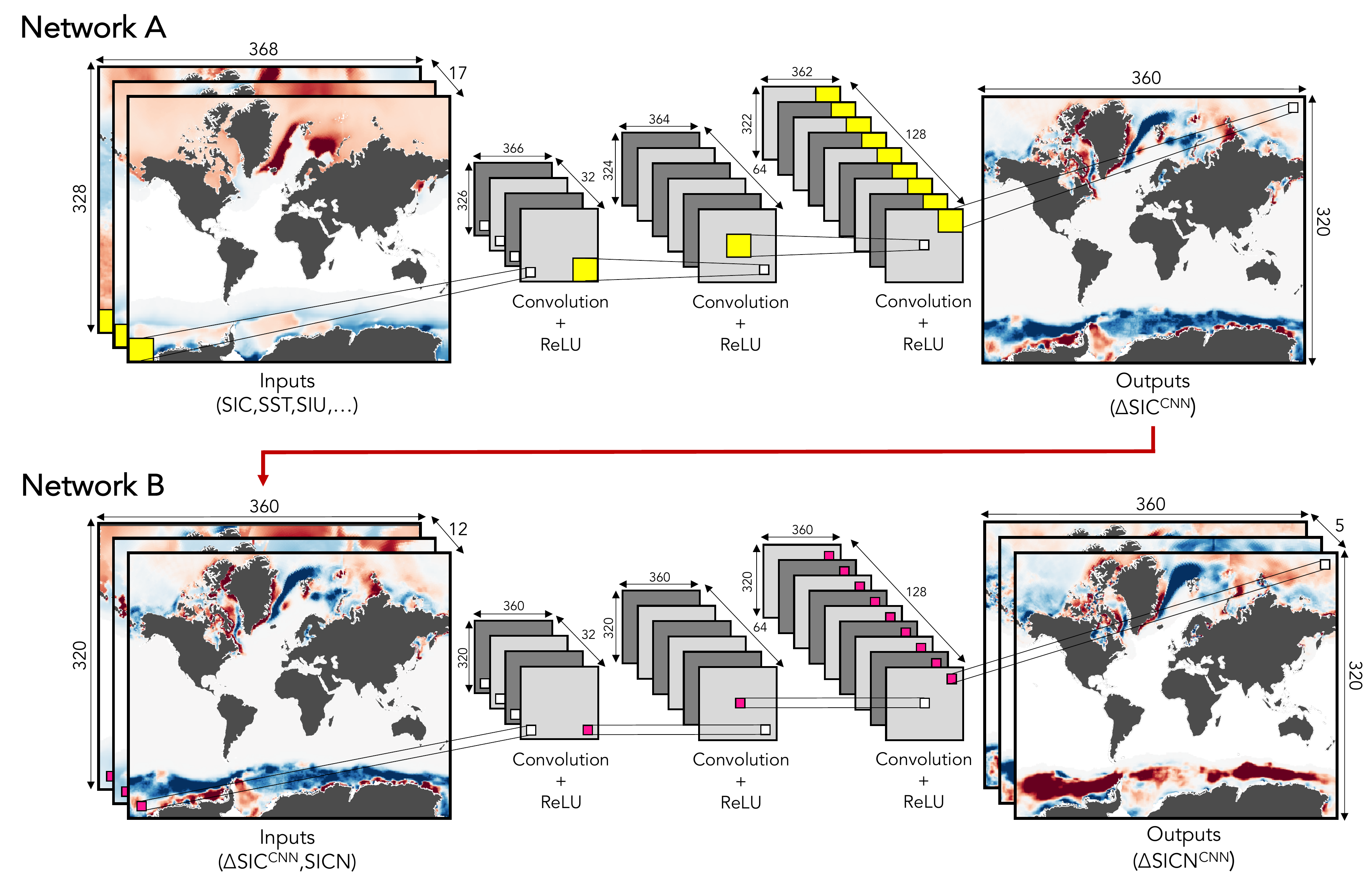}
    \caption{Schematic of the CNN architectures used to learn functional mappings from state vectors to analysis increments. The yellow and purple squares represent 3$\times$3 and 1$\times$1 kernels over which the convolution operations are performed in each layer, respectively, where there is one kernel for every feature map in each layer. The white pixel is then the sum of convolution outputs from all features in the previous layer, which has subsequently been passed through a ReLU activation function. The activation function after the last convolution operation to the output layer is the identity function.}
    \label{fig:CNN}
\end{figure}

\section{Convolutional neural networks}\label{sect:CNNs}
CNNs are a specific class of DL algorithms which are well-suited to problems where inputs contain local correlation structure in space and/or time \cite{Murphy2022}. For this reason they have historically been successful in the domains of image recognition and segmentation \cite{Simonyan2014,Zeiler2014,Dong2015,Ronneberger2015,Krizhevsky2017}, where the aim is to e.g., classify objects or isolate features within medical images. In Earth system modeling CNNs have subsequently been utilized for their ability to exploit the two-dimensional structure associated with turbulent fluids, and hence learn subgrid parameterizations of ocean mesoscale eddies \cite{Bolton2019,Zanna2020} and cloud moisture convection \cite{Han2020}. For this reason, we use them here to learn sea ice model errors, which also inherently exhibit two-dimensional structure.

\begin{table}[t!]
\center
\caption{Details of the convolutional neural networks (CNNs) inputs, outputs, architecture, and hyperparameters used during training.}
\begin{tabular}{p{4.5cm} p{3cm} p{3cm}} & \textbf{Network A} & \textbf{Network B} \\
\hline
\textbf{Inputs} (* states \& tendencies) & SIC*, SST*, SIU*, SIV*, SIT*, SW*, TS*, SSS*, Land-sea mask & $\Delta$SIC\textsuperscript{CNN}, SICN*, Land-sea mask \\
\textbf{Outputs} & $\Delta$SIC & $\Delta$SICN \\
\textbf{Size of input data set} & $2094 \times 17 \times 328 \times 368$ & $2094 \times 12 \times 320 \times 360$ \\
\textbf{Size of output data set} & $2094 \times 1 \times 320 \times 360$ & $2094 \times 5 \times 320 \times 360$ \\
\textbf{Normalization} & Inputs standardized (see main text) & Inputs standardized (see main text) \\
\textbf{Convolution layers}  & 4 & 4 \\
\textbf{Features per layer}  & 32, 64, 128, 1 & 32, 64, 128, 5 \\
\textbf{Activation function(s)} & ReLU, ReLU, ReLU, Linear & ReLU, ReLU, ReLU, Linear \\
\textbf{Kernel size(s)} & $3 \times 3$ & $1 \times 1$ \\
\textbf{Kernel stride(s)} & 1 & 1\\
\textbf{Bias parameters} & False & False\\
\textbf{Zero-padding} & None & None \\
\textbf{Total weights} & 98,208 & 11,264 \\
\textbf{Batch size} & 10 & 10 \\
\textbf{Optimizer} & Adam & Adam \\ 
\textbf{Learning rate} & 0.001 & 0.001 \\
\textbf{Weight decay} & $1 \times 10^{-7}$ & $1 \times 10^{-7}$ \\
\textbf{Epochs} & 150 & 125 \\
\textbf{Seed} & 711 & 711 \\
\end{tabular}
\label{tab:tab1}
\end{table}

\subsection{Architecture}\label{sect:architecture}
Generally speaking, a CNN can be seen as a series of linear weighted sums in which a rectangular weight matrix, or \emph{kernel}, slides over an input image in order to produce a new feature representation of that same input. By sequentially repeating this procedure on each new feature map, and adding nonlinear activation functions between network layers, the network is then able to extract increasingly complex behaviour from the inputs, before a final operation which maps the last set of features to each pixel of the output image. Figure \ref{fig:CNN} shows this procedure in the present context of learning `images' of sea ice DA increments. In this case we develop two independent CNNs, where each can be classified as a `fully CNN' as the outputs of each layer are produced only by convolution operations. Network A is used to learn the aggregate ($\Delta$SIC) increments from various atmosphere, ocean and sea ice model states and forecast tendencies, while network B uses the predictions of $\Delta$SIC from network A in order to learn a mapping from $\Delta$SIC to $\Delta$SICN. We find this two-step approach yields significantly lower prediction error than using a single network to predict $\Delta$SICN directly. Table \ref{tab:tab1} summarizes the architectural choices made for Networks A and B.

Each of the inputs of a given CNN have independent kernels that connect to every feature map in the subsequent layer of the network, hence with $3\times3$ kernels in each layer, 17 input variables, and features per layer of 32, 64, 128, and 1, network A has a total number of weights given by $(3 \times 3 \times 17 \times 32) + (3 \times 3 \times 32 \times 64) + (3 \times 3 \times 64 \times 128) + (3 \times 3 \times 128 \times 1) = 98,208$. Meanwhile, with 1$\times$1 kernels in each layer, 12 input variables, and features per layer of 32, 64, 128, and 5, network B has a total number of weights given by $(1 \times 1 \times 12 \times 32) + (1 \times 1 \times 32 \times 64) + (1 \times 1 \times 64 \times 128) + (1 \times 1 \times 128 \times 5) = 11,264$. An advantage of the CNN approach is that a single kernel matrix is used for the entire spatial domain of a given input, meaning that structures which exhibit similar characteristics, but occur at different locations within the input, will be equally resolved. This property of \emph{translational invariance} is not guaranteed in e.g., typical feed-forward (artificial) neural networks which use the whole domain at once as input \cite{Gardner1998}. Non-linearities within the system can also be exploited by passing each feature map through a non-linear activation function, such as the rectified linear unit (ReLU) function, which is the identity function for positive values and zero for negative values. In both networks in our application, the first three convolution operations are followed by ReLU activation functions, while the final convolution to the output layer is simply linear.

The inputs to network A correspond to the 5-day means of the model states and 5-day forecast tendencies from each DA cycle, for each of SIC, sea-surface temperature (SST), zonal and meridional components of ice velocities (SIU and SIV, respectively), sea ice thickness (SIT), net shortwave radiation (SW), ice-surface skin temperature (TS), sea-surface salinity (SSS), and finally a land-sea mask containing zeros over land grid cells and ones over ocean grid cells. Note that SIU and SIV are vector fields with values located at C-grid cell edges, while the other scalar fields have values centered within each grid cell (see e.g., \citeA{Griffies2004}). This means that SIU and SIV contain one additional matrix column and row, respectively, compared to the scalar fields. We therefore compute a 2-point average along the columns of SIU and rows of SIV, so that the these variables are defined on the same tracer grid as the scalar fields. The inputs to network B correspond to the $\Delta$SIC predictions from network A, along with the model states and forecast tendencies of SICN, as well as a land-sea mask. It should also be noted that the inputs of each network (excluding the land-sea mask) are standardized by subtracting their respective mean and normalizing by their respective standard deviation, where both mean and standard deviations are computed over ocean grid cells poleward of 40$^\circ$ latitude, across all training samples (see section 
\ref{sect:train}). This provides a single value of the mean and standard deviation for each network input. Furthermore, given that, in our network architecture, each convolution operation in network A reduces the size of the input image by 2 pixels in both matrix dimensions, the final outputs are 8 pixels smaller than the original inputs (hence a 9$\times$9 centered stencil is required to make a local prediction at any grid point). To ensure we utilize the appropriate information at the image boundaries, we therefore pad the input data by 4 pixels on each side in the following way: the last 4 columns of the image are padded in front of the first column (zonal periodicity), the original first 4 columns are padded to the last column (zonal periodicity), a copy of the first 4 rows is flipped 180$^\circ$ counter-clockwise and padded in front of the first row (symmetry across the model's Arctic bipolar fold, see \citeA{Griffies2004}; the sign of the ice velocities in the first 4 rows is also flipped during this process), and finally the last row is padded with 4 rows of zeros (the final row corresponds to the Antarctic continental land mass). 

\subsection{Training}\label{sect:train}
In order to generate accurate predictions, the weights of each CNN must be optimized. This is typically achieved by minimizing an appropriate loss function $\mathcal{L}$ which describes the similarity between the final outputs of the network and the target variable (i.e., the analysis increments). For network A the loss function ($\mathcal{L}_A$) is the mean-squared error (MSE) of the $\Delta$SIC predictions, while for network B the loss function ($\mathcal{L}_B$) is the sum of the MSE of each of the five $\Delta$SICNs, as well as an additional term to impose a soft constraint that the sum of the five $\Delta$SICNs are equal to $\Delta$SIC:

\begin{flalign}
    \hspace{-118pt} \mathcal{L}_A = \frac{1}{NS}\sum_{i=1}^{NS} \big(\Delta\mathrm{SIC}^{\mathrm{CNN}}_i - \Delta\mathrm{SIC}^{\mathrm{True}}_i\big)^2,
\label{SICerror}
\end{flalign}

\begin{equation}
\begin{aligned}
    \mathcal{L}_B &= \sum_{k=1}^5 \frac{1}{NS} \sum_{i=1}^{NS} \big(\Delta\mathrm{SICN}^{\mathrm{CNN}}_{k_i} - \Delta\mathrm{SICN}^{\mathrm{True}}_{k_i}\big)^2 \\
     & \hspace{50pt}  + \lambda\bigg(\frac{1}{NS}\sum_{i=1}^{NS} \big(\sum_{k=1}^5 \Delta\mathrm{SICN}^{\mathrm{CNN}}_{k_i} - \sum_{k=1}^5 \Delta\mathrm{SICN}^{\mathrm{True}}_{k_i}\big)^2\bigg).
\label{SICNerror}
\end{aligned}
\end{equation}

Here, $N = 320 \times 360 = 115,200$ is the number of model grid points, which corresponds to the entire globe. $S = 10$ is the batch size (randomly shuffled temporal samples), and $\lambda = 5$ is a scaling constant. The loss function is minimized using the Adam stochastic gradient descent method \cite{Kingma2014} within the PyTorch Python library \cite{Paszke2019}, which accommodates graphical processing unit (GPU) and batch processing facilities for significant computational speed-ups and efficient memory handling, respectively. Recall Table \ref{tab:tab1} for a full list of the details of each CNN.

As well as optimizing the weights of each CNN, there are other factors which influence the predictive performance that also need to be considered. For one, there is the physical architecture of each CNN, which includes e.g., the number of layers within each network, the type of activation function, and the size of the convolution kernels. Then there are also specific hyperparameters, which include e.g., the learning rate of the Adam optimizer, and the number of training epochs. Choosing the optimal architectures and hyperparameters is referred to as \emph{model selection} and is generally approached by selecting the model which produces the lowest error score on unseen validation data (i.e., data that were not used to optimize the CNN weights). In order to ensure that the validation error is representative of the model's predictive performance across all samples it is often necessary to perform $K$-fold cross-validation, where the data are split into $K$ equal-sized temporally contiguous chunks. The model is then trained on $K-1$ chunks, and predictions are validated on the remaining chunk. We opt for temporally contiguous chunks here, as opposed to random sampling of training and validation points, due to inherent temporal auto-correlation within the data, which would likely lead to data leakage issues during the validation stage. In any case, this process is repeated $K$ number of times where each time a different chunk is chosen to be the validation set. The average validation error across all $K$ tests is then the generalization error of that particular CNN model. To arrive at the final CNN architectures and hyperparameters detailed in Table \ref{tab:tab1}, we performed 5-fold cross-validation at each model selection step, hence for a given architecture and set of hyperparameters the model was trained 5 times, where each time the 2618 temporal samples were split into different combinations of 2094 training and 524 validation points. Specific architectures and hyperparameters were subsequently chosen based on the model which showed the lowest average 5-fold cross-validation score. Ideally, one would perform model selection by scanning all possible combinations of hyperparameters and CNN architectures and finding which combination produces the lowest cross-validation score. For large data sets however, this is computationally impractical and as such we proceeded with model selection by testing one hyperparameter and/or architecture at a time and taking the model with the lowest 5-fold cross-validation score forward to the next test (see Figure \ref{fig:S3} for example learning curves from various model selection tests). The results in the next section are based on predictions on validation data from the final CNN models, as described in Table \ref{tab:tab1}. Note that, for convenience, hereafter we refer to networks A and B together as our final network architecture.

\section{Results}\label{sect:Results}
Before presenting the results of the CNN predictions, we first introduce the error metrics which are used to evaluate the model's performance. For a given spatial map of the SIC increments on any given day, $\Delta$SIC\textsuperscript{True}, and the equivalent CNN prediction on the same day, $\Delta$SIC\textsuperscript{CNN}, the regional uncentered spatial pattern correlation \cite{Barnett1987} between these two fields is given as:
\begin{equation}
    \rho = \frac{\sum_{i=1}^n \Delta\mathrm{SIC}_i^\mathrm{CNN}\Delta\mathrm{SIC}_i^\mathrm{True}}{\Vert \Delta\mathrm{SIC}^\mathrm{CNN}\Vert_2\Vert \Delta\mathrm{SIC}^\mathrm{True}\Vert_2},
\label{SPC}
\end{equation}
where $\Vert \cdot \Vert_2$ is the $\ell_2$ vector norm, and $n = 100 \times 360 = 36,000$ for either pan-Arctic or pan-Antarctic regions (approx. 45$^\circ$N and 30$^\circ$S, respectively). We opt for this metric over the standard (centered) linear correlation coefficient as the subtraction of the mean to compute the covariance in the centered case may result in differences between $\Delta$SIC\textsuperscript{True} and $\Delta$SIC\textsuperscript{CNN} at open-ocean grid cells (e.g., \citeA{Legates1997}). Similar to the centered pattern correlation, an uncentered pattern correlation value of 1 represents a perfect agreement between the true and predicted increments on day $t$, while a value of $-1$ represents a perfect out-of-phase agreement. A value of 0 subsequently represents no agreement.

We also introduce the regional root-MSE (RMSE) as:
\begin{equation}
    \mathrm{RMSE} = \sqrt{\frac{1}{n} \sum_{i=1}^{n}\big(\Delta\mathrm{SIC}_i^\mathrm{CNN} - \Delta\mathrm{SIC}_i^\mathrm{True}\big)^2}.
\label{RMSE}
\end{equation}
This metric captures the average deviation of the predictions from the true increments, hence an RMSE value of 0 represents perfect predictions.

\begin{figure}[t!]
    \centering
    \includegraphics[width=1\linewidth]{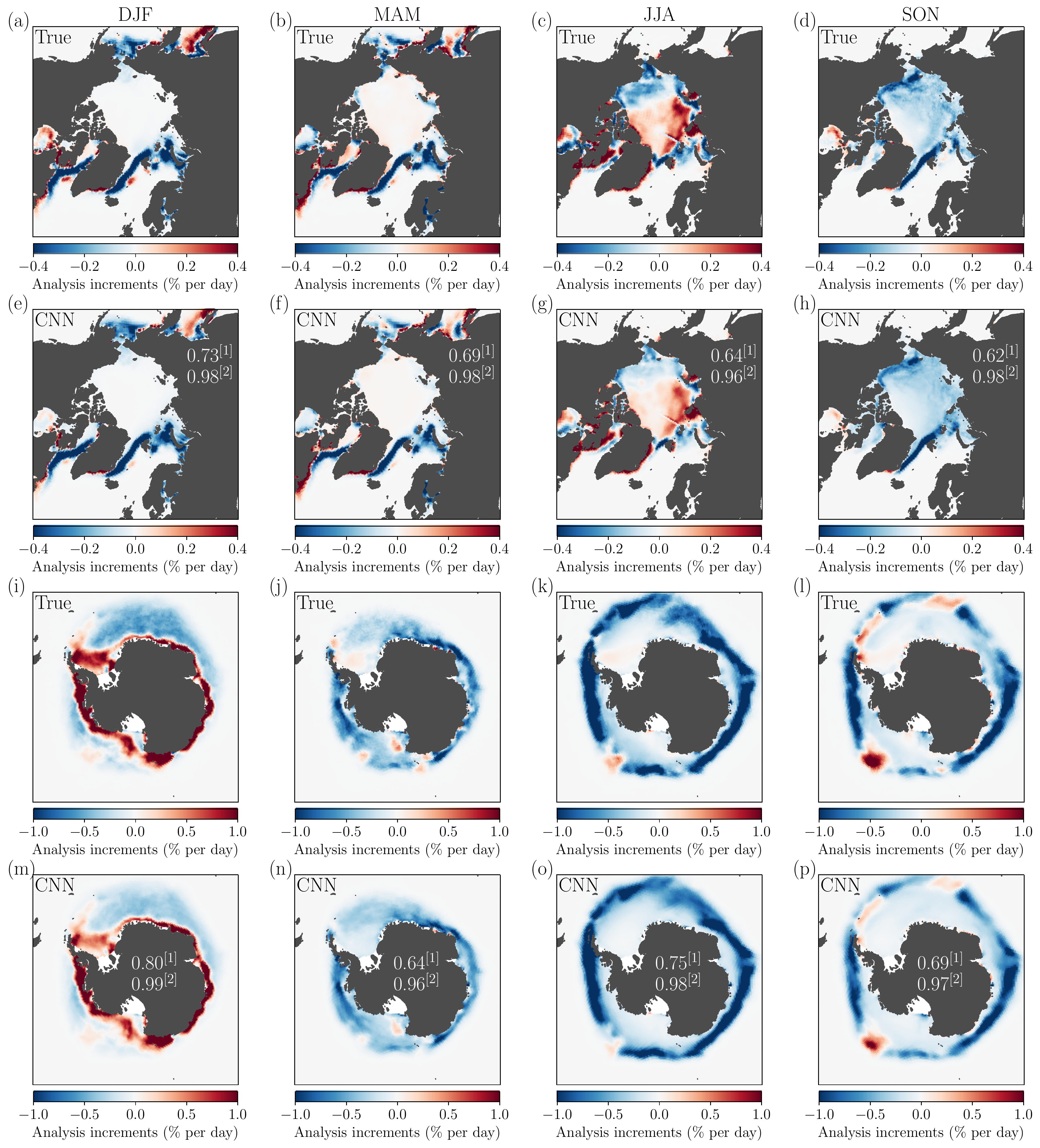}
    \caption{Seasonal climatologies of the (true) SPEAR aggregate sea ice concentration analysis increments and the equivalent CNN predictions, for both the Arctic (a)--(h) and Antarctic (i)--(p). Columns from left to right show DJF, MAM, JJA, SON climatologies, computed over the period 1982--2017. Values with the superscript [1] are the average of daily spatial pattern correlations between $\Delta$SIC\textsuperscript{True} and $\Delta$SIC\textsuperscript{CNN} in each respective season, while values with [2] are the spatial pattern correlations between the respective climatologies of the true and predicted increments.}
    \label{fig:avepredictions}
\end{figure}

\begin{figure}[t!]
    \centering
    \includegraphics[width=1\linewidth]{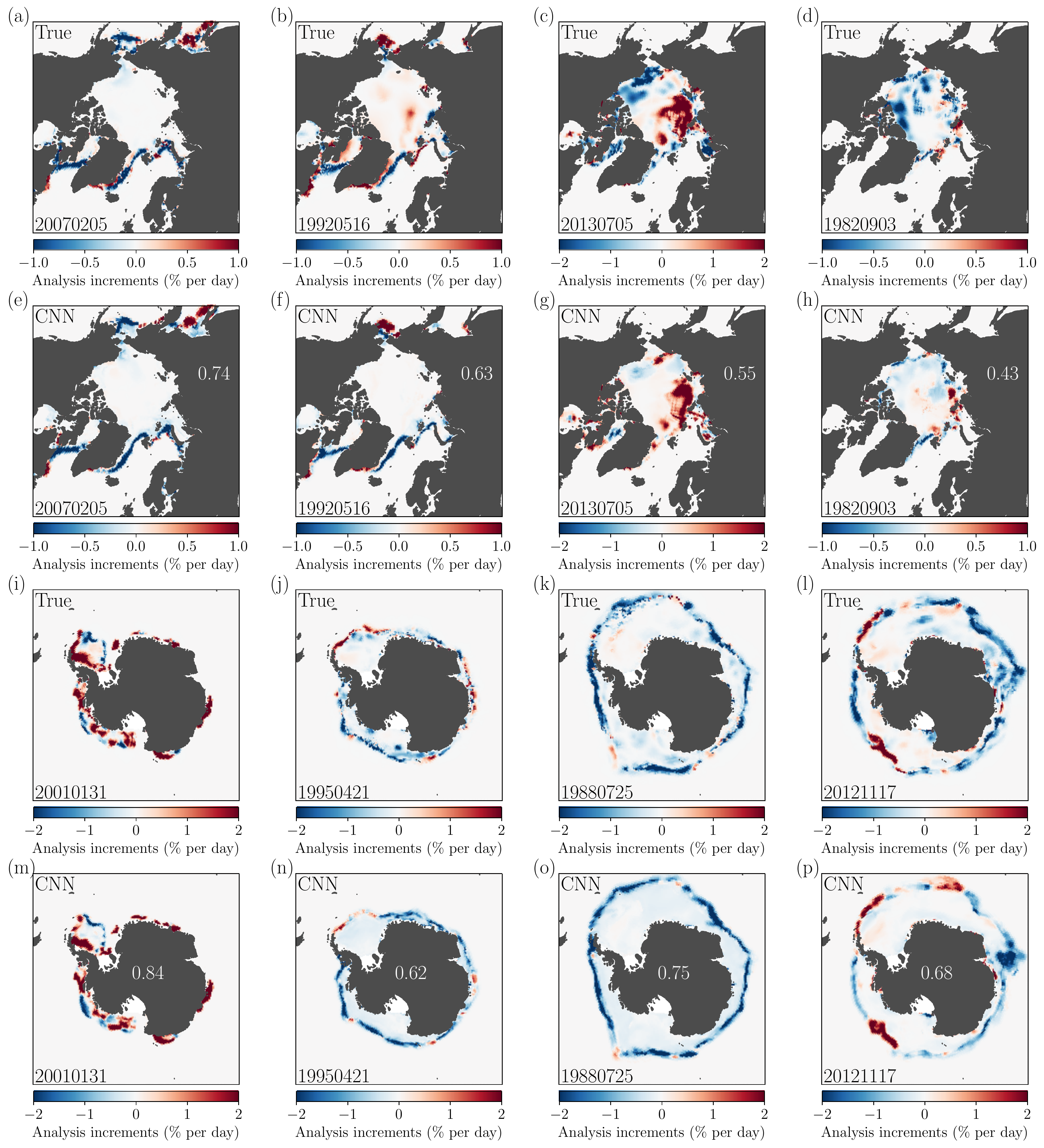}
    \caption{Daily snapshots of the (true) SPEAR aggregate sea ice concentration analysis increments and the equivalent CNN predictions, for both the Arctic (a)--(h) and Antarctic (i)--(p). Columns from left to right show random days in DJF, MAM, JJA, and SON over the period 1982--2017. Spatial pattern correlations are reported for each prediction.}
    \label{fig:snapshots}
\end{figure}

\subsection{Predictions}
In this section we show the predictions of $\Delta$SIC as the sum of the five predicted $\Delta$SICNs, on the held-out data that were not used to optimize the network weights during training. We therefore generate 2618 predictions spanning the 1982--2017 period, which correspond to combining the 5 individual held-out chunks from the cross-validation experiment of the final model, into a continuous time series record. We focus on $\Delta$SIC here, as opposed to $\Delta$SICN, as the former is the direct observable quantity and as such lends to more intuitive interpretation of the results, although the reader is referred to Figures \ref{fig:S4}--\ref{fig:S8} for comparable versions of Figure \ref{fig:avepredictions} for each $\Delta$SICN.

Figure \ref{fig:avepredictions} shows the seasonal climatologies of the $\Delta$SIC predictions, where we notice that, in both hemispheres, the CNN is able to predict the average spatial pattern of the increments very well. In the Arctic, the network performs best in DJF, with average daily spatial pattern correlations of 0.73, and a spatial pattern correlation of 0.98 between the climatologies of the daily DJF predicted and true increments. The poorest predictions in the Arctic are in JJA and SON with average daily spatial pattern correlations of 0.64 and 0.62, respectively, and correlations of 0.96 and 0.98, respectively between the climatologies. In JJA for example, while the network reproduces the average spatial pattern well, the magnitude of the increments to the north of Greenland and in the Canada basin is generally too low. Similarly, in the Antarctic, the CNN also performs best in DJF with average daily spatial pattern correlations of 0.80, however the average magnitude of the predicted increments is generally too low in regions such as the Weddell Sea. The poorest predictions in the Antarctic are in MAM with average daily spatial pattern correlations of 0.64, perhaps owing to the network's inability to fully resolve the relatively small-scale heterogeneities in e.g., the Ross and Weddell seas. The large-scale patterns are generally in good accordance however. These initial results suggest that the network is able to learn the mean bias patterns of the model with considerable skill.

Moving beyond assessments of climatologies, Figure \ref{fig:snapshots} shows randomly sampled snapshots of the predictions for individual days across each season, as a way to assess how the CNN performs at capturing the fast physics errors (for an animation of the CNN performance on additional daily snapshots, see Supporting Information S2). Broadly speaking, we find that the CNN is able to capture the large-scale structure of the increments, but often fails to capture smaller-scale features. The February prediction in the Arctic (Figure \ref{fig:snapshots}e) shows high skill with a spatial pattern correlation of 0.74, however at this time of year the increments are primarily associated with sea ice edge errors, while the increments in the central ice pack (i.e., the majority of the Arctic domain) are effectively zero. Nonetheless, the CNN is able to predict these ice edge errors very well, particularly in the Labrador, GIN, Barents, Okhotsk, and Bering seas. As the melt season progresses, the prediction skill generally drops, where it is lowest in September (Figure \ref{fig:snapshots}h), with a spatial pattern correlation of 0.43. The true July and September increments (Figures \ref{fig:snapshots}c and \ref{fig:snapshots}d, respectively) exhibit significant variability within the core ice pack which, in some regions, the network is unable to reproduce. For example, the large negative increments in the Beaufort and Chukchi seas in July. The CNN does however manage to capture some amount of the variability in July, such as the large positive increments in the Kara and Laptev shelf seas. 

The prediction skill in the Antarctic is generally higher than in the Arctic, and comparing Figures \ref{fig:snapshots}i and \ref{fig:snapshots}m, we can see that the CNN accurately predicts a significant amount of the variability in summer, with a spatial pattern correlation of 0.84. The subsequent predictions in April, July and November (Figures \ref{fig:snapshots}n-p) show slightly lower skill than in January, with the lowest skill in April with a spatial pattern correlation of 0.62. At these times the increments are largely related to sea ice edge errors, and the CNN is generally able to capture the large-scale patterns, as well as some of the localized features, such as the positive increments at the north-eastern edge of the Ross Sea in November (Figure \ref{fig:snapshots}p), and the band of positive increments along the northern edge of the Weddell Sea in April (Figure \ref{fig:snapshots}n).

From the daily snapshots we can infer that the CNN captures large amounts of the fast physics errors, although there is some seasonal variation to the skill, where the predictions in the Arctic are generally best over the winter period and poorest in the summer. Meanwhile in the Antarctic the predictions appear most skillful in the summer and poorest in the early growth season (April). In the next section we provide an assessment of the CNN's sensitivity to various inputs, as well as its sensitivity to the geographic training domain. In doing so, we subsequently highlight this seasonal skill variation in more detail.

\begin{figure}[t!]
    \centering
    \includegraphics[width=1\linewidth]{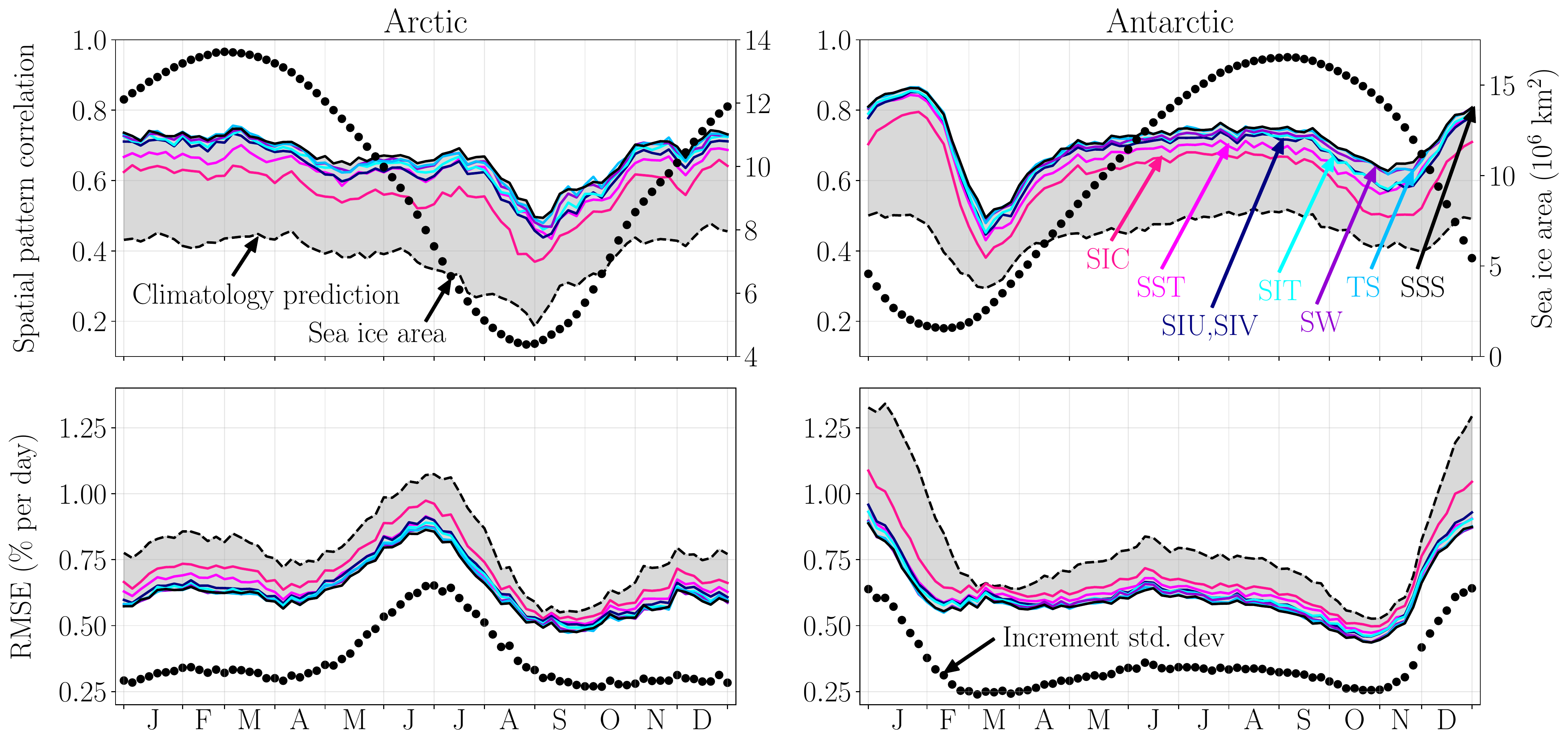}
    \caption{Prediction skill metrics for independent sensitivity tests to network inputs, presented as daily climatologies of predictions on held-out samples, computed over the period 1982--2017, for the Arctic (left column) and Antarctic (right column). The shaded region reflects the improvement in skill of the final network (solid black curve) over the benchmark climatology prediction (dashed black curve).}
    \label{fig:hierarchy}
\end{figure}

\subsection{Sensitivity analysis}
\subsubsection{Network inputs}
In this section we perform sensitivity tests to determine which model states and forecast tendencies contribute most to the prediction skill of $\Delta$SIC (again, as the sum of the five predicted $\Delta$SICNs on held-out samples), at different times of the year. The sensitivity analysis is performed by training a series of initial networks which each contain a single variable as inputs (e.g., SIC states and forecast tendencies), and assessing which of these networks results in the highest prediction skill of $\Delta$SIC in both hemispheres. The input variable of this network is then assumed to be the most physically-relevant predictor of $\Delta$SIC. The testing then continues by training a second series of networks which contain two input variables: the best predictor from the first test, as well as any one of the remaining input variables. The network which results in the largest improvement in skill relative to the best network from the first test is then taken forward, and so on. For 7 network input variables (classifying SIU and SIV as a single input), we therefore trained 28 independent network configurations in order to establish a hierarchy of predictors.

Figure \ref{fig:hierarchy} shows daily 36-year climatologies of spatial pattern correlation and RMSE error metrics, for sensitivity tests in both the Arctic and Antarctic domains. The hierarchy of predictors in terms of largest skill contribution proceeds as: SIC, SST, SIU and SIV, SIT, SW, TS, and finally SSS. Hence for the SIC curves, the network inputs to generate these predictions are only SIC states and forecast tendencies, while for the SST curves, the network inputs are SIC and SST states and forecast tendencies, and so on. The SSS curve then represents the predictions from the final model (i.e., the network architecture presented in section \ref{sect:architecture}). The climatology prediction (black dashed curve) refers to using the daily 36-year climatology of the true $\Delta$SIC increments to predict the true $\Delta$SIC increment on any given day. This is an offline-equivalent to the `ocean tendency adjustment' approach by \citeA{Lu2020}, as discussed in section \ref{sect:intro}, and as such serves as our benchmark here, where we can see that each sensitivity test provides improvement in skill over this climatological tendency benchmark. From this analysis we can also see that, relative to the benchmark climatology, SIC is responsible for a significant fraction of the overall network skill (approx. 66\% in both hemispheres). SST, SIU and SIV then account for an additional 20\%, with the remaining variables SIT, SW, TS and SSS making up the last 14\%. Furthermore, while SIC, SST, SIU and SIV are essential inputs in all months of the year, the contributions from other variables such as SW and TS are generally limited to the summer months.

In terms of spatial pattern correlation, the maximum skill of the final network in the Arctic occurs at the beginning of March, after which the skill declines somewhat continuously until the end of July, and then  more rapidly to its minimum in early September. Meanwhile in the Antarctic, the points of maximum and minimum skill are separated by approximately 1.5 months, with the maximum occurring at the end of January, and the minimum at the beginning of March. Although the skill variation in the Arctic appears to somewhat correlate with the climatological sea ice area, the rate of change in sea ice area in the melt and growth season is generally not consistent with that of the CNN prediction skill. Furthermore, in the Antarctic the skill is increasing between November and January, while the sea ice area is decreasing. This therefore suggests that the skill variation is not directly tied to the seasonal cycle of sea ice area. When also considering the standard deviation of the increments, we can see that the low spatial pattern correlation scores coincide with times when the standard deviation of the increments, and hence the RMSE, are relatively low. This may initially suggest that the lower spatial pattern correlation at these times is either a consequence of low signal variance, or that the network training does little to optimize these points as they inherently have lower MSE than e.g., the winter months. If however the spatial pattern correlation scores were a direct reflection of the increment standard deviation, we would expect to see similarly low spatial pattern correlations in e.g., April in the Arctic or November in the Antarctic, however this is not generally the case. What is noticeable, is that the climatology benchmark also exhibits the same seasonal variation in spatial pattern correlation and RMSE as the CNN predictions, highlighting that the lower skill in the late summer in both hemispheres is not likely due to any shortcomings in the ML model, but rather a feature of the increments themselves. In particular, the climatological prediction is less skillful in the low-CNN-skill months, suggesting that these months are inherently more challenging to predict. 

\begin{figure}[t!]
    \centering
    \includegraphics[width=1\linewidth]{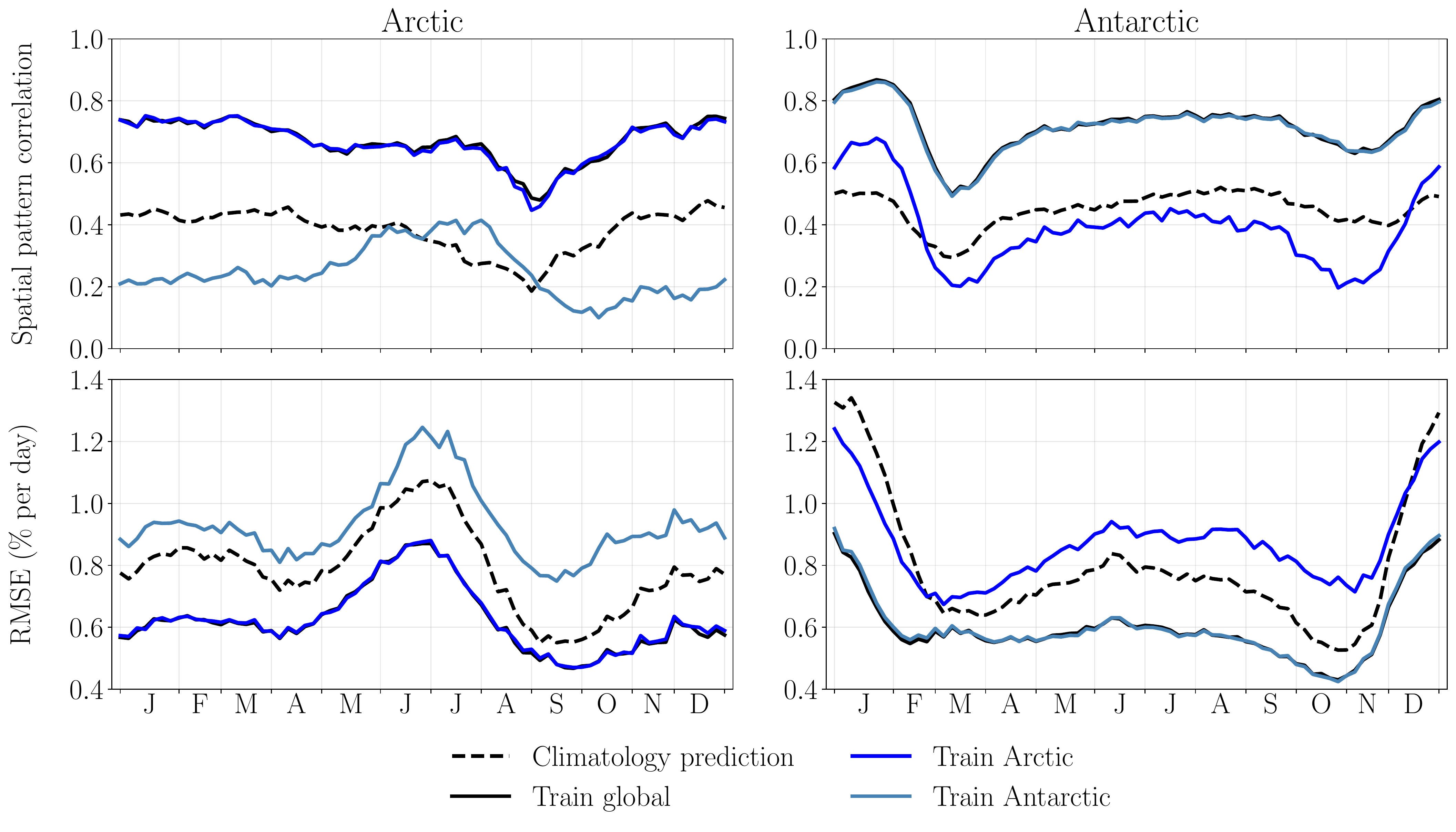}
    \caption{Prediction skill metrics for independent sensitivity tests to the network training domain, presented as daily climatologies of predictions on held-out samples, computed over the period 1982--2017, for the Arctic (left column) and Antarctic (right column).}
    \label{fig:domain}
\end{figure}

\subsubsection{Training domain}
The network in this study is trained on data from the entire globe, meaning that it must find the optimal set of weights which generalize to make accurate predictions of the analysis increments in both the Arctic and the Antarctic. Given that the bias patterns, and hence characteristics of the increments, are somewhat different between the two hemispheres, we conduct further sensitivity tests to determine how well the network has generalized. As before, error metrics are shown in terms of the sum of the five predicted $\Delta$SICNs on held-out samples that were not used to train the model.

Figure \ref{fig:domain} shows daily climatologies of spatial pattern correlation and RMSE error metrics, for three variations of the network training setup. One where the network is trained on the entire globe (i.e., our proposed network in section \ref{sect:architecture}), one where the network is trained on just the Arctic domain, and one where the network is trained on just the Antarctic domain. Here we notice that the network which is trained on global data is able to make just as skillful predictions of $\Delta$SIC in the Arctic, as the network which is trained only on Arctic data. The same is also true for the Antarctic case. Interestingly, we can also see that the network which is trained only on the Arctic are still able to make relatively skillful predictions of the Antarctic increments, even performing better than the benchmark climatology predictions between the months of December and February. Meanwhile, the network which is trained on Antarctic data is not able to generalize as well to the Arctic, although still shows some small amount of skill between July and August. This analysis therefore confirms that training on global data is vital for generalizing across domains while still matching the skill of networks trained on each individual domain.

\begin{figure}[t!]
    \centering
    \includegraphics[width=1\linewidth]{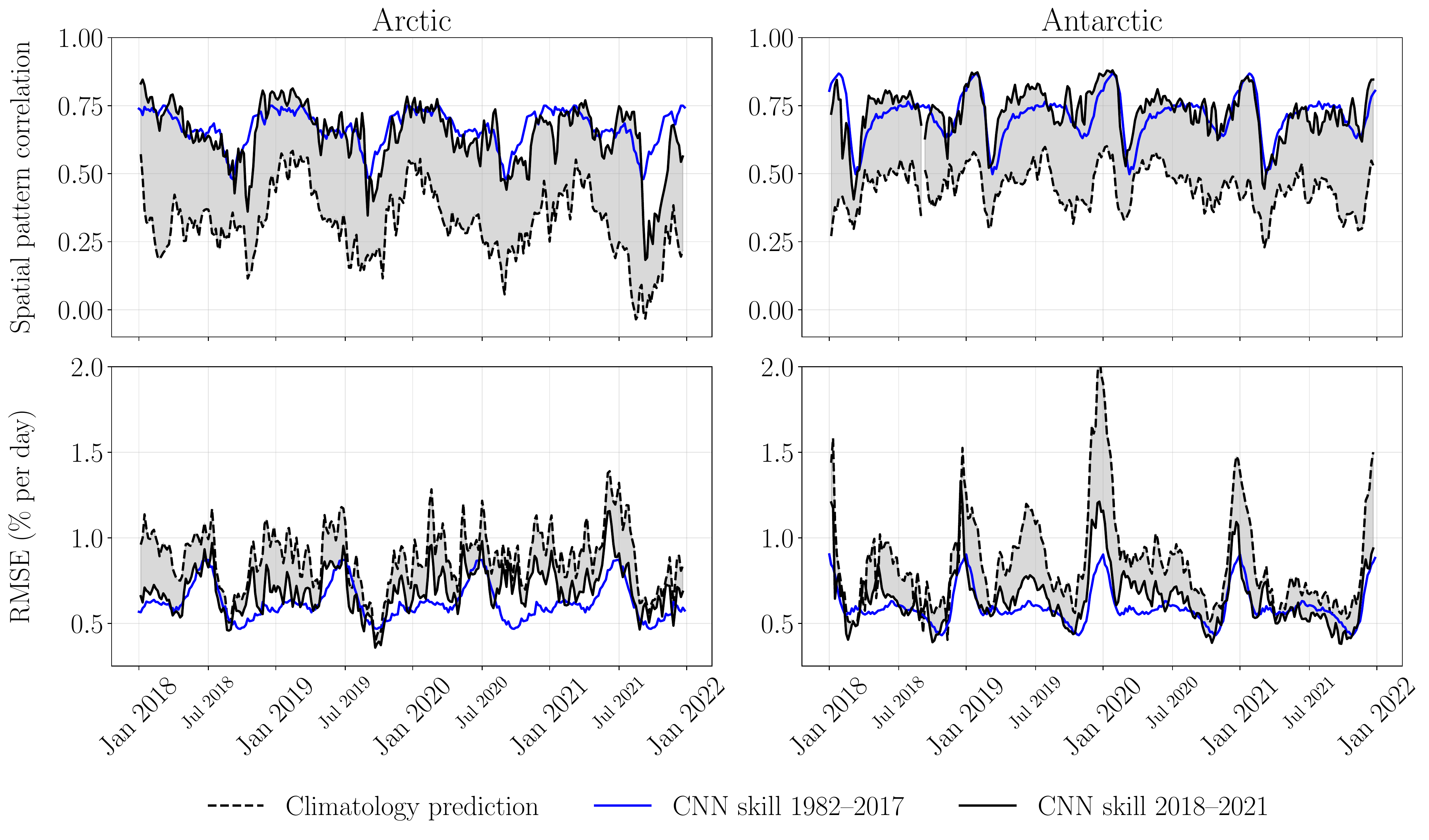}
    \caption{Generalization performance of CNN predictions for the extended period between January 2018 and December 2021. Error metrics for the black curves are shown at the frequency of the data assimilation system (5-daily), while the blue curve is the daily climatology skill of the final network over the 1982--2017 period.}
    \label{fig:generalization}
\end{figure}

\subsection{Final validation}
Due to the fact that we perform model selection by choosing specific CNN architectures and hyperparameters which minimize the average cross-validation score on data that were not used to optimize the CNN weights, there is an inherent risk of over-fitting the model to these validation data. As such, it is often necessary to retain an additional data set which has not been used for validation at any point during the model selection process. For this, we extend the \citeA{Zhang2021} sea ice DA experiment from December 27\textsuperscript{th} 2017, through to December 27\textsuperscript{th} 2021, providing an additional 291 validation data points. We subsequently evaluate the performance of our CNN model by training on all 2618 samples between 1982--2017, and validating on the extended data period between 2018--2021. It should be noted that this extended DA experiment is identical in configuration to that which was outlined in section \ref{sect:DA}, except that in this extended case the atmospheric forcing from JRA55-do reanalysis corresponds to version 1.5, while previously it was version 1.3. This version change relates to a correction in the sign and rotation of tropical cyclones, and as such we do not expect this to result in significant differences in the representation of sea ice in the extended DA simulations. 

Figure \ref{fig:generalization} shows daily spatial pattern correlation and RMSE error metrics over the 2018--2021 period for both the Arctic and Antarctic domains (black curves). We also overlay the daily climatology skill of the final network architecture from the cross-validation experiments between 1982--2017, hence the blue curves here are identical to the `Train global' curves in Figure \ref{fig:domain}, and are simply repeated for each of the 4 validation years presented. The predictions appear to generalize well to the future data, where spatial pattern correlation values are generally in accordance with the 1982--2017 period, particularly in the Antarctic, and are still out-performing the climatology prediction in both hemispheres. On average, the RMSE over the 2018--2021 period is slightly higher than the 1982--2017 climatology, which is due to the fact that there is a non-stationary component to the increments, whereby the variance increases over the course of the time series record (see Figure \ref{fig:S9}). Therefore naturally the climatological RMSE of the CNN predictions increases over time as well (see Figure \ref{fig:S10}). In any case, the generalization ability of the predictions suggests that the CNN has not simply over-fitted to the training and/or validation data during model selection.

\section{Discussion}\label{sect:Disc}
The ability of the proposed CNN to make skillful predictions of the sea ice concentration analysis increments, using only information on local model state variables and their tendencies, provides interesting avenues for future work. The fact that the predictions show improvements in skill relative to a daily increment climatology (e.g., \citeA{Lu2020}), generalize well to each hemisphere, and show skill on a separate validation data set, strongly suggests that the CNN could be used to reduce sea ice biases within SPEAR, either as an online sea ice model parameterization, or as a bias correction tool for numerical sea ice prediction. Ultimately, one could argue that there is still room for improvement in the CNN performance, particularly in the late summer months. Considering the inherent complexity of the problem at hand, and the likely influence of both non-linear and non-local processes, it is conceivable to push the limit of predictive skill further by increasing the complexity of the network, both in terms of the total number of weights, and the 9$\times$9 grid cell domain of influence on a local prediction. Indeed, such changes could be implemented through increasing the width and/or depth of the network, as well as incorporating non-local connections (in space) through e.g., fully-connected layers. On the other hand, the architectures here been developed specifically with the goal of a sea ice model parameterization in mind, and as such, factors including computational cost and practicality of implementation in parallelized high-performance computing environments have been considered throughout the development. In the following sections we provide a discussion on the directions for future work relating to both sea ice parameterization and seasonal sea ice prediction. 

\subsection{Considerations for parameterization}
ML models have been shown to be successful at parameterizing subgrid-scale processes within dynamical models, including ocean mesoscale eddies \cite{Guillaumin2021}, atmospheric convection \cite{Yuval2020}, and sea ice dynamics \cite{Finn2023}. Common to each of these studies is that the ML models target specific physical processes, with the aim of replacing pre-existing knowledge-based parameterizations, or deriving new parameterizations for physical processes which are not currently represented. On the other hand, our proposed CNN is trained to predict sea ice increments which reflect numerous interacting model errors across various model components. To subsequently disentangle these coupled model physics errors a-posteriori and then apply them as parameterizations to their respective components, is not straightforward. In our goal of constructing a sea ice model parameterization, it is critical to ensure that the parameterization is not acting to correct coupled model errors that originate in other model components (e.g., an ocean heat transport bias or atmospheric circulation bias that imprints upon the sea ice). Our DA-ML methodology attempts to mitigate this possibility, as the ice-ocean DA system is driven by atmospheric reanalysis and also nudges SST and SSS towards observed values. These observational constraints on the atmosphere and ocean allow us to interpret the DA increments as isolated sea ice model physics errors, however this assumption is not perfect as the ocean component of the DA system can still imprint some errors on the sea ice state (e.g., Figure \ref{fig:S1}). Future investigation will be required to determine how the CNN generalizes in a fully coupled setting with fully-interactive atmosphere-ice-ocean feedbacks (see section \ref{sect:forecasting}).

Another major consideration for a sea ice parameterization is how to appropriately conserve mass, heat, and salt. In the context of the ocean, \citeA{Lu2020} achieved global conservation of heat and salinity when implementing the climatological ocean DA increments into MOM6 by ensuring that the global integral of the correction to each variable was zero. In the case of sea ice, assuming the parameterization enters the thermodynamic solver, then appropriately coupling this parameterization with the upper ocean would mean that a predicted negative sea ice concentration increment would remove sea ice (column-wise) by adding mass and salt to the ocean mixed layer, while also removing heat. This step is likely to come in the form of a mass, heat, and salt budget assessment between the ice and ocean after evaluating the amount of local sea ice mass change associated with a given predicted SIC increment, rather than adapting the CNN architectures themselves to respect conservation. 

Although we have considered implementation cost in the design of our network, some investigation will be required to quantify this cost in terms of both matrix computations and additional memory load. Regarding memory load, our parameterization will not require any additional memory in terms of the number of grid cells stored on any one central processing unit (CPU), as our 9$\times$9 network stencil requires the same number of `halo' grid points as the default SPEAR configuration, which uses a halo size of 4. There will be some small amount of memory cost for storing the network weights on each CPU however. Looking to similar studies, \citeA{Guillaumin2021} found that implementing a fully CNN with 8 convolutional layers as a stochastic parameterization into an idealized shallow water model resulted in a 25\% increase in the run time, compared to an unparameterized simulation. \citeA{Zhang2023} also found that the cost of doing inference with this same network as a parameterization in MOM6 was 10 times more expensive than the CPU cost of the simulation itself. Although we effectively have 8 convolutional layers when considering both networks A and B, we can still expect much lower computational overheads given that ours is a deterministic model (i.e., we predict a single output at each grid point for each $\Delta$SICN, rather a, potentially larger, number of parameters which describe a distribution of values), and that our kernel size for network B is 1$\times$1 in each layer, while the \citeA{Guillaumin2021} network uses variable size kernels throughout, ranging from sizes 3$\times$3 to 5$\times$5. Like in this study, they also did not use zero-padding, though in their case given the larger kernel sizes, they required a stencil of 21$\times$21 grid points to make a local prediction. 

Finally, the increments in this study represent error growth over a 5-day period, and the input states and tendencies of the CNN are given as 5-day means. After implementation, the CNN predictions will need to produce a correction which reflects error growth over a given model timestep, and similarly the input states and tendencies will need to be adjusted accordingly. This will therefore require further sensitivity tests to determine how to appropriately perform this scaling.

\subsection{Considerations for sea ice forecasting}\label{sect:forecasting}
Some of the initial concerns over implementation of the CNN as a sea ice model parameterization can be alleviated by assessing how the network performs as an online bias correction tool within the context of seasonal sea ice forecasting. In previous work, \citeA{Zhang2022} showcased the benefits of using SIC assimilation to initialize the sea ice conditions for SPEAR retrospective forecasts (hereafter reforecasts) of the Arctic sea ice cover between 1992--2017. In \citeA{Zhang2022}, the same ice-ocean SPEAR model configuration and initial conditions as outlined here in section \ref{sect:DA}, were used to perform DA between 1982 and the first day of each month, for all years between 1992 and 2017. Whereby the first day of each month represented the initialization point, after which the model would run in fully coupled mode to generate forecasts out to 1-year lead time. Assimilation in \citeA{Zhang2021,Zhang2022} was performed by passing the prior model state variables and observations to the Data Assimilation Research Testbed (DART; \citeA{Anderson2009}), which then computes the set of analysis states offline, providing the new set of initial conditions with which to begin the next assimilation cycle. Given that our CNN is inherently independent of the observations, we propose that it would be relatively straightforward to bias correct the sea ice within the fully coupled reforecast period by replacing the standard call to DART with our CNN. In this scenario, we could perform seasonal reforecasts (up to 12-month lead times) to assess how the network generalizes to the fully coupled SPEAR model, while not requiring strict conservation properties due to the shorter time scales. Furthermore, we could continue in the same 5-day cycle configuration so that the network predictions would not need to be scaled for different temporal sampling. If the reforecasts then have improved skill relative to the SPEAR DA-initialized reforecasts from \citeA{Zhang2022}, they may be fit-for-purpose as a model parameterization.  

\section{Concluding remarks}\label{sect:Conc}
\subsection{Summary}
In this study we have shown that deep learning (DL), specifically convolutional neural networks (CNNs), can be used to make skillful predictions of sea ice model errors, in the form of data assimilation (DA) increments, using only information from model state variables and tendencies (the time derivative of the model state variables). We developed a CNN using an ice-ocean DA system which assimilates satellite observations of sea ice concentration (SIC) into the Seamless system for Prediction and EArth system Research (SPEAR) model every 5 days between 1982--2017. SPEAR has a 5-category ice thickness distribution, hence concentration increments are produced for each subgrid category, where the observable (aggregate) SIC increment corresponds to the sum of 5 categories. We therefore developed a two-step CNN architecture, in which the first step learns the physical mapping from various local sea ice, ocean and atmosphere state variables and forecast tendencies to the aggregate SIC increments. The second step then learns the mapping from the aggregate concentration error to each of the subgrid terms. We subsequently showed that our DL architecture is able to make skillful predictions of the SIC increments in both the Arctic and the Antarctic and across all seasons. Spatial pattern correlations between the climatologies of the observed and predicted increments are high, with values of at least 0.96 for both the Arctic and Antarctic, demonstrating that the CNN is able to skillfully capture the mean model bias. The CNN also has skill at predicting the state-dependent model errors, with daily pattern correlation values ranging from 0.64--0.80 and 0.62--0.73 in the Antarctic and Arctic, respectively. This shows that the CNN is able to predict the fast physics errors and systematic bias patterns of the SPEAR model with considerable skill, which is also confirmed by the fact that the predictions show improved skill over a model which simply predicts the climatological mean increment on any given day of the year. Sensitivity analysis revealed that SIC as an input to the network is responsible for approximately 66\% of the overall network skill, followed by sea-surface temperature (SST) and ice velocities which account for 20\%, and finally ice thickness, net shortwave radiation, ice-surface skin temperature and sea-surface salinity which account for the remaining 14\%. 

\subsection{Outlook}
Recent studies have highlighted how DA provides a unique opportunity to leverage sparse and/or noisy observations, in order to facilitate machine learning of structural model errors \cite{Bonavita2020,Brajard2021,Farchi2021,Mojgani2022,Chen2022}. Building on this, we have shown here how DA also provides the ability to learn errors within unobserved model state variables, and hence provides a new framework for learning subgrid-scale parameterizations for climate models. In section \ref{sect:Disc} we subsequently outlined how the strong predictive performance of the CNN and its generalization ability suggests that the network has the potential to reduce sea ice biases in free-running climate simulations, as a sea ice model parameterization within SPEAR. Irrespective of this eventual goal however, the findings in this work ultimately have wider implications for the climate modeling and numerical weather prediction (NWP) community in general. With regards to NWP, previous studies have already shown that ML techniques can be used to learn state-dependent fast physics errors within large-scale atmospheric models, subsequently leading to improved online predictions by using the ML model as a bias correction tool \cite{Bonavita2020,Chen2022}. In our study, we have shown that the concept of learning state-dependent fast physics errors is transferable to a global ice-ocean model, which could further aid NWP when considering that coupling the atmosphere with an ice-ocean model has previously shown to improve short-term weather predictions \cite{Smith2018}.

Turning to longer-term simulations, the fact that the systematic errors are also predictable suggests that a parameterization built from DA increments has the potential to reduce persistent climate model biases and improve the fidelity of climate change projections. On the other hand, while we have shown that state variables such as SIC and SST explain a significant fraction of the variance in the analysis increments, our current framework does not allow us to attribute these correlations to a specific model deficiency, for example an incorrectly parameterized or missing physical process. One additional avenue for future work could therefore involve designing a perfect model experiment in which a single ensemble member is run with a specific parameterization that has been tuned or turned on (e.g., sea ice ridging or melt-pond formation). This member would then be treated as the ground truth and assimilated into the original model. The resultant analysis increments would then be a manifestation of the systematic bias within the original model, associated with this specific incorrect/missing parameterization, and hence one could more confidently isolate which state variables within an ML model contribute most to predicting this particular structural error. 

\section{Open Research} 
\noindent All data for training each CNN are openly available at the following locations:
\begin{itemize}
\item Inputs (DA forecast states and tendencies): \url{ftp://sftp.gfdl.noaa.gov/perm/William.Gregory/seaice_DA-ML_inputs_1982-2017.nc}
\item Outputs (DA increments): \url{ftp://sftp.gfdl.noaa.gov/perm/William.Gregory/seaice_DA-ML_outputs_1982-2017.nc}
\end{itemize}
Python code to pre-process the input data and train the CNNs can also be found at \url{https://github.com/m2lines/seaice_DA-ML}. The optimized weights of the CNNs and standardization statistics for the inputs are also saved within the same repository.

\acknowledgments
William Gregory, Mitchell Bushuk, Alistair Adcroft and Laure Zanna received M$^2$LInES research funding by the generosity of Eric and Wendy Schmidt by recommendation of the Schmidt Futures program. This work was also intellectually supported by various other members of the M$^2$LInES project, as well as being supported through the provisions of computational resources from the National Oceanic and Atmospheric Administration (NOAA) Geophysical Fluid Dynamics Laboratory (GFDL). We also thank Spencer Clark and Zachary Labe for their invaluable feedback on this article.


%
%



\bibliography{main}

\newpage
\setcounter{figure}{0} 
\renewcommand{\thefigure}{S\arabic{figure}}
\title{Supporting Information S1: Deep learning of systematic sea ice model errors from data assimilation increments}

\authors{William Gregory\affil{1}, Mitchell Bushuk\affil{2}, Alistair Adcroft\affil{1}, Yongfei Zhang\affil{1}, Laure Zanna\affil{3}}

\affiliation{1}{Atmospheric and Oceanic Sciences Program, Princeton University, NJ, USA}
\affiliation{2}{Geophysical Fluid Dynamics Laboratory, NOAA, Princeton, NJ, USA}
\affiliation{3}{Courant Institute of Mathematical Sciences, New York University, New York, NY, USA}

\correspondingauthor{Will Gregory}{wg4031@princeton.edu}

\vspace{50pt} 

\begin{figure}[h!]
    \centering
    \includegraphics[width=1\linewidth]{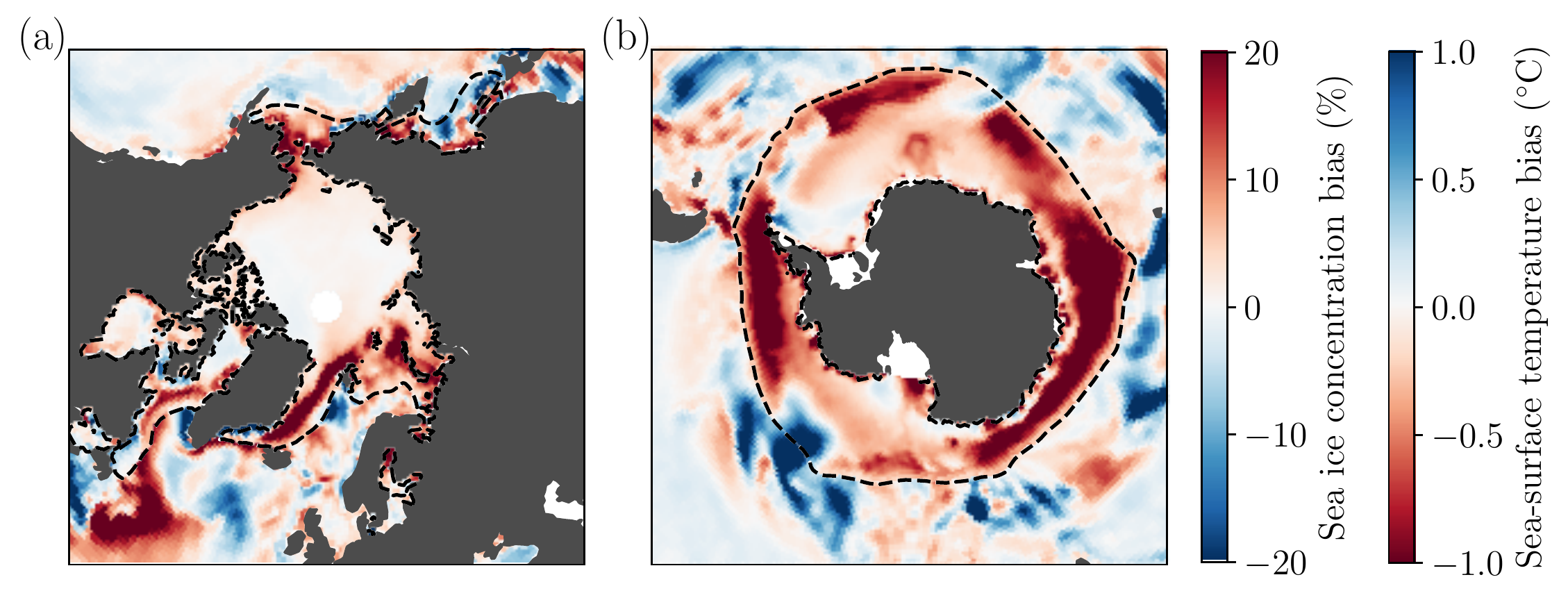}
    \caption{Seasonal climatologies of the SPEAR free-running model bias. Inside the climatological sea ice extent contour (black dashed line) are the aggregate sea ice concentration biases (model minus observations). Outside the contour are the sea-surface temperature biases (model minus observations). a) Arctic biases (December--February), b) Antarctic biases (September--November).}
\label{fig:S1}
\end{figure}

\begin{figure}[t!]
    \centering
    \includegraphics[width=.75\linewidth]{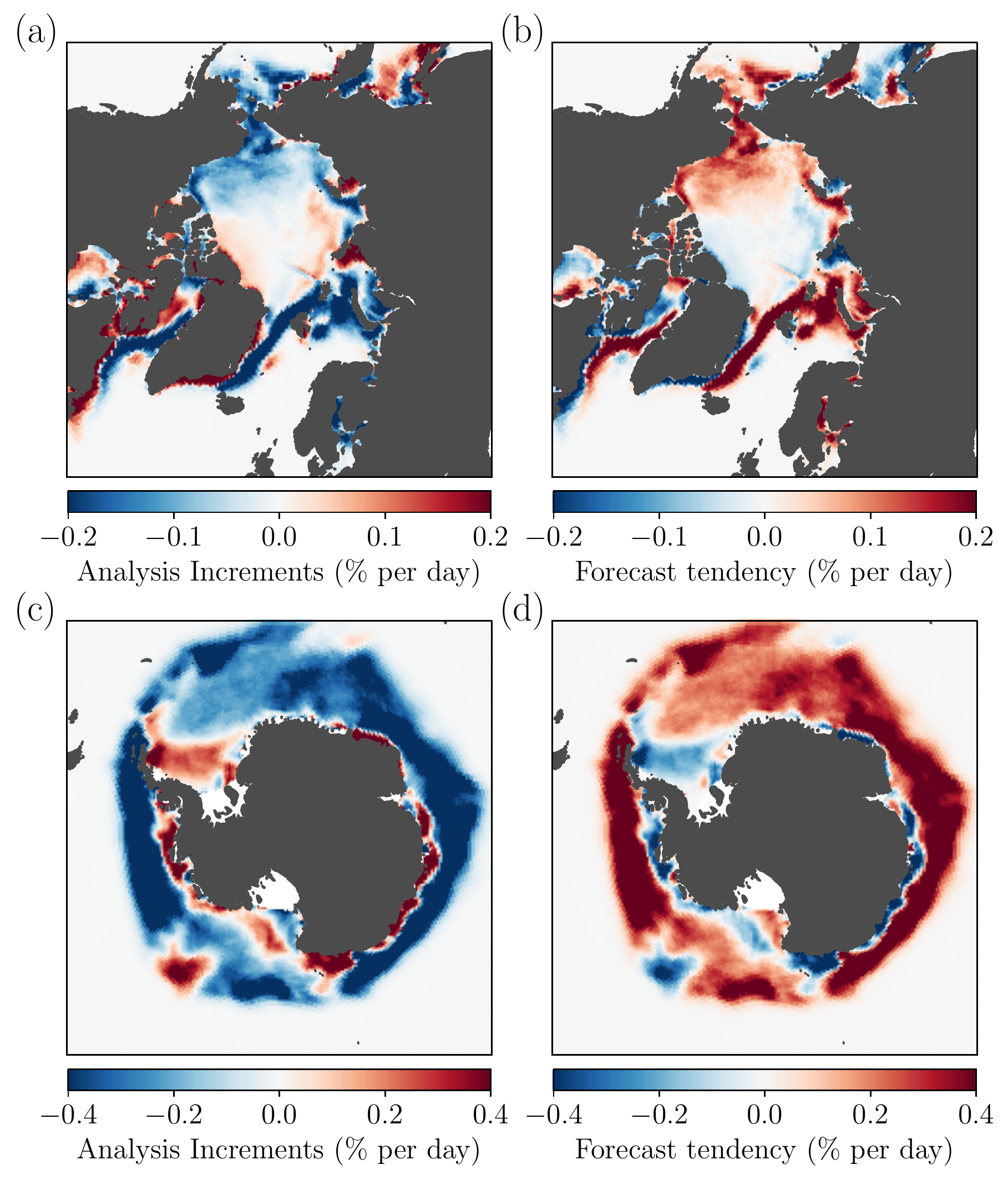}
    \caption{Comparisons of the climatological aggregate sea ice concentration analysis increments (a,c), and the aggregate sea ice concentration forecast tendencies (b,d). The spatial pattern correlation between panels a) and b) is -0.99. Similarly, the spatial pattern correlation between panels c) and d) is also -0.99.}
\label{fig:S2}
\end{figure}

\begin{figure}[t!]
    \centering
    \includegraphics[width=\linewidth]{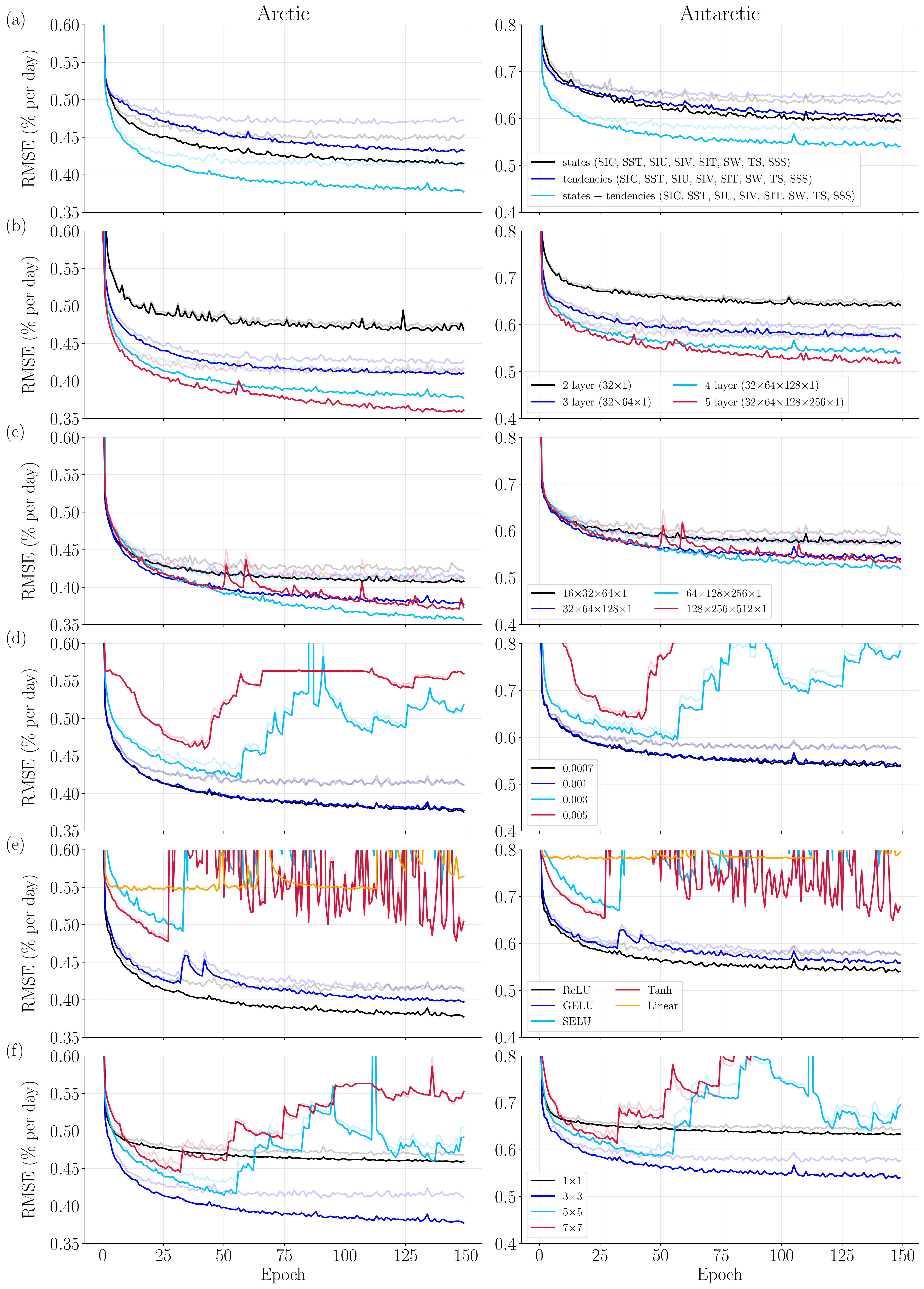}
    \caption{Learning curve examples for various CNN model selection tests. Each curve is the mean 5-fold cross-validation error on $\Delta$SIC predictions (solid lines = error on training samples, transparent curves =  error on validation samples). (a) Tests of the network sensitivity to the inputs (i.e., using just state variables, or just tendencies, or both). (b) Tests of the network depth (number of convolutional layers). (c) Tests of the network width (features per convolutional layer). (d) Tests of the optimizer learning rate. (e) Tests of the activation function used after each convolution operation. (f) Tests of the size of the convolution kernel used in each layer.}
\label{fig:S3}
\end{figure}

\begin{figure}[t!]
    \centering
    \includegraphics[width=1\linewidth]{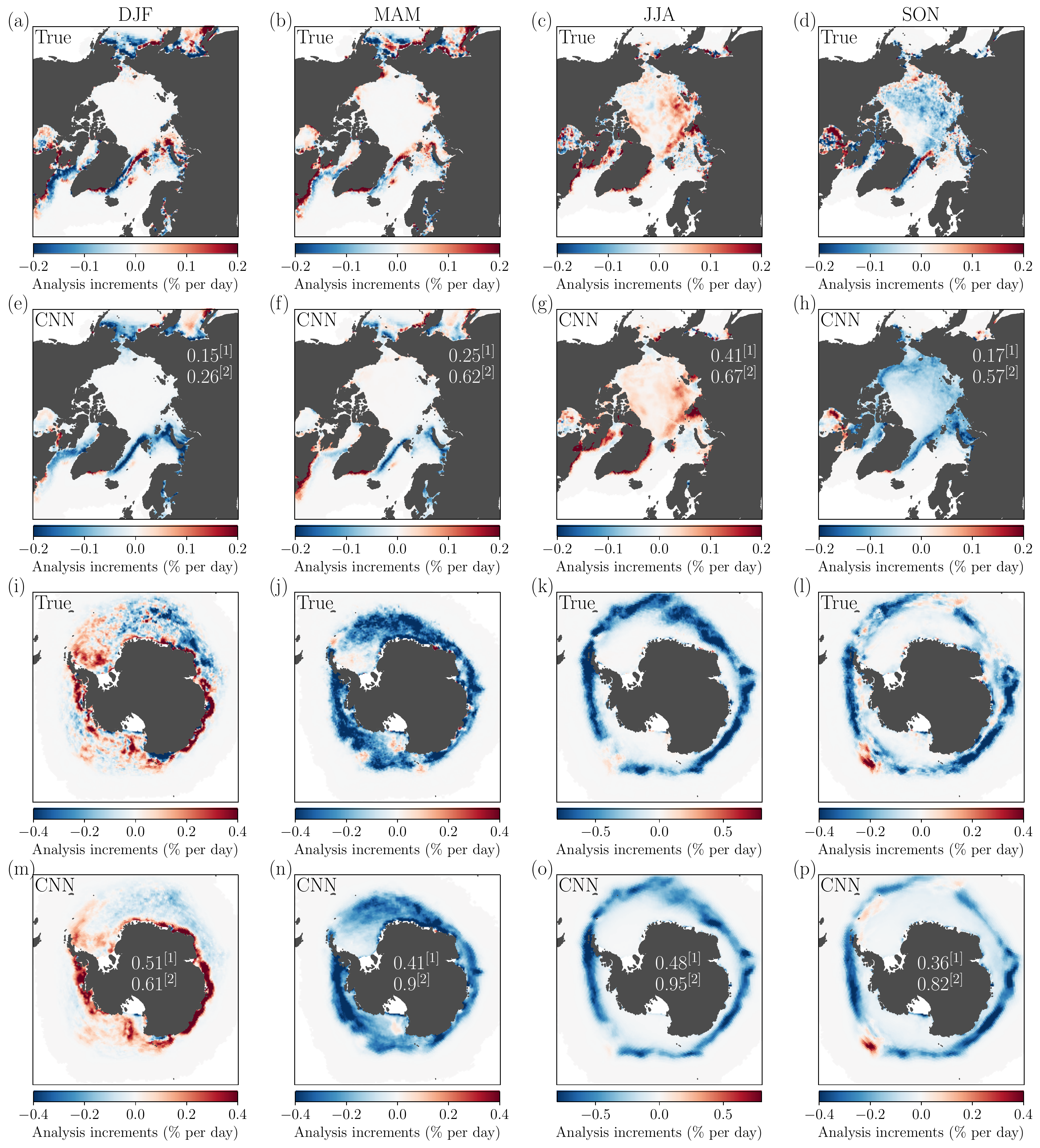}
    \caption{Seasonal climatologies of the (true) SPEAR category 1 sea ice concentration analysis increments and the equivalent CNN predictions, for both the Arctic (a)--(h) and Antarctic (i)--(p). Columns from left to right show DJF, MAM, JJA, SON climatologies, computed over the period 1982--2017. Values with the superscript [1] are the average of daily spatial pattern correlations between $\Delta$SICN\textsuperscript{True} and $\Delta$SICN\textsuperscript{CNN} in each respective season, while values with [2] are the spatial pattern correlations between the respective climatologies of the true and predicted increments.}
\label{fig:S4}
\end{figure}

\begin{figure}[t!]
    \centering
    \includegraphics[width=1\linewidth]{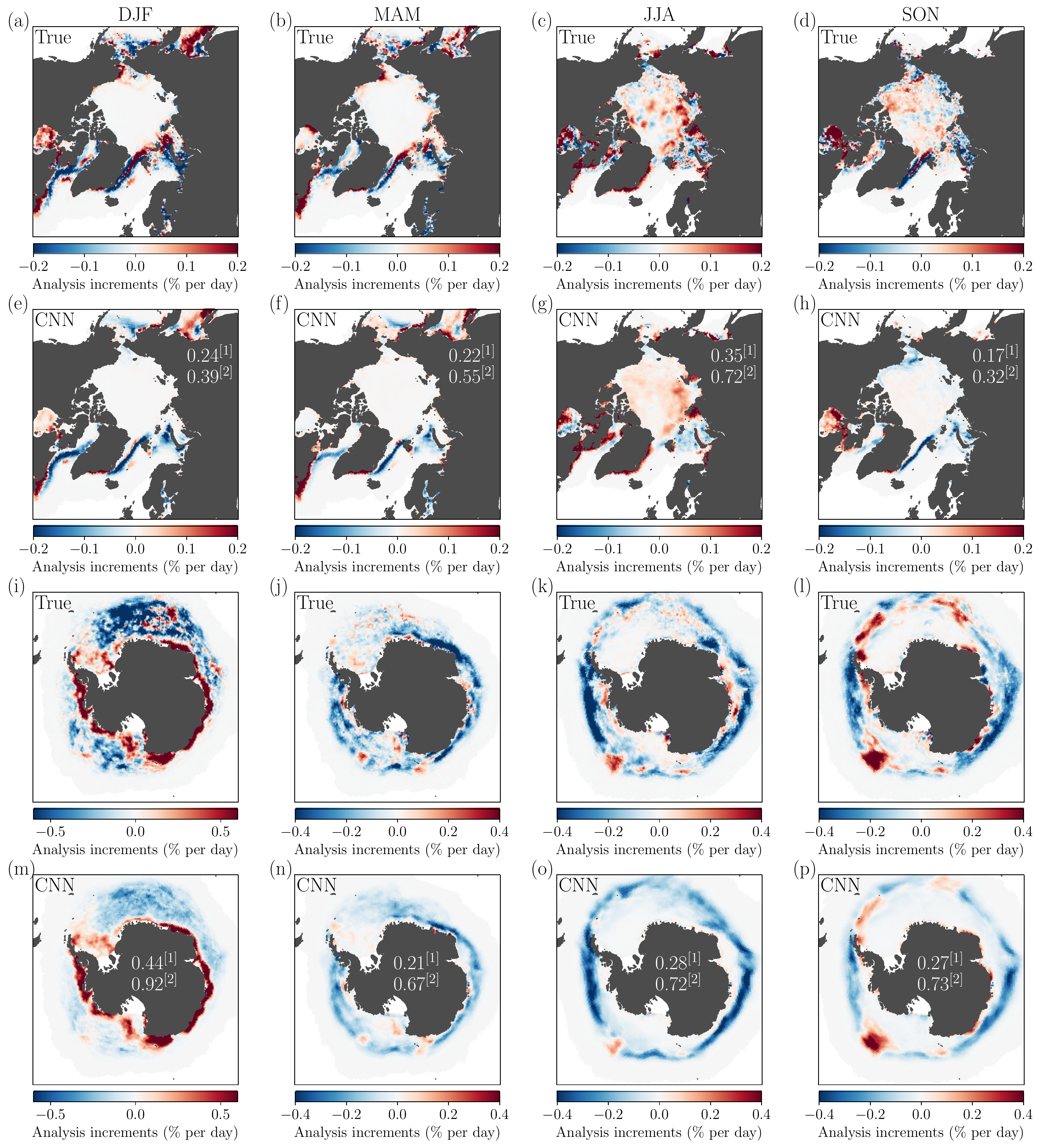}
    \caption{Seasonal climatologies of the (true) SPEAR category 2 sea ice concentration analysis increments and the equivalent CNN predictions, for both the Arctic (a)--(h) and Antarctic (i)--(p). Columns from left to right show DJF, MAM, JJA, SON climatologies, computed over the period 1982--2017. Values with the superscript [1] are the average of daily spatial pattern correlations between $\Delta$SICN\textsuperscript{True} and $\Delta$SICN\textsuperscript{CNN} in each respective season, while values with [2] are the spatial pattern correlations between the respective climatologies of the true and predicted increments.}
\label{fig:S5}
\end{figure}

\begin{figure}[t!]
    \centering
    \includegraphics[width=1\linewidth]{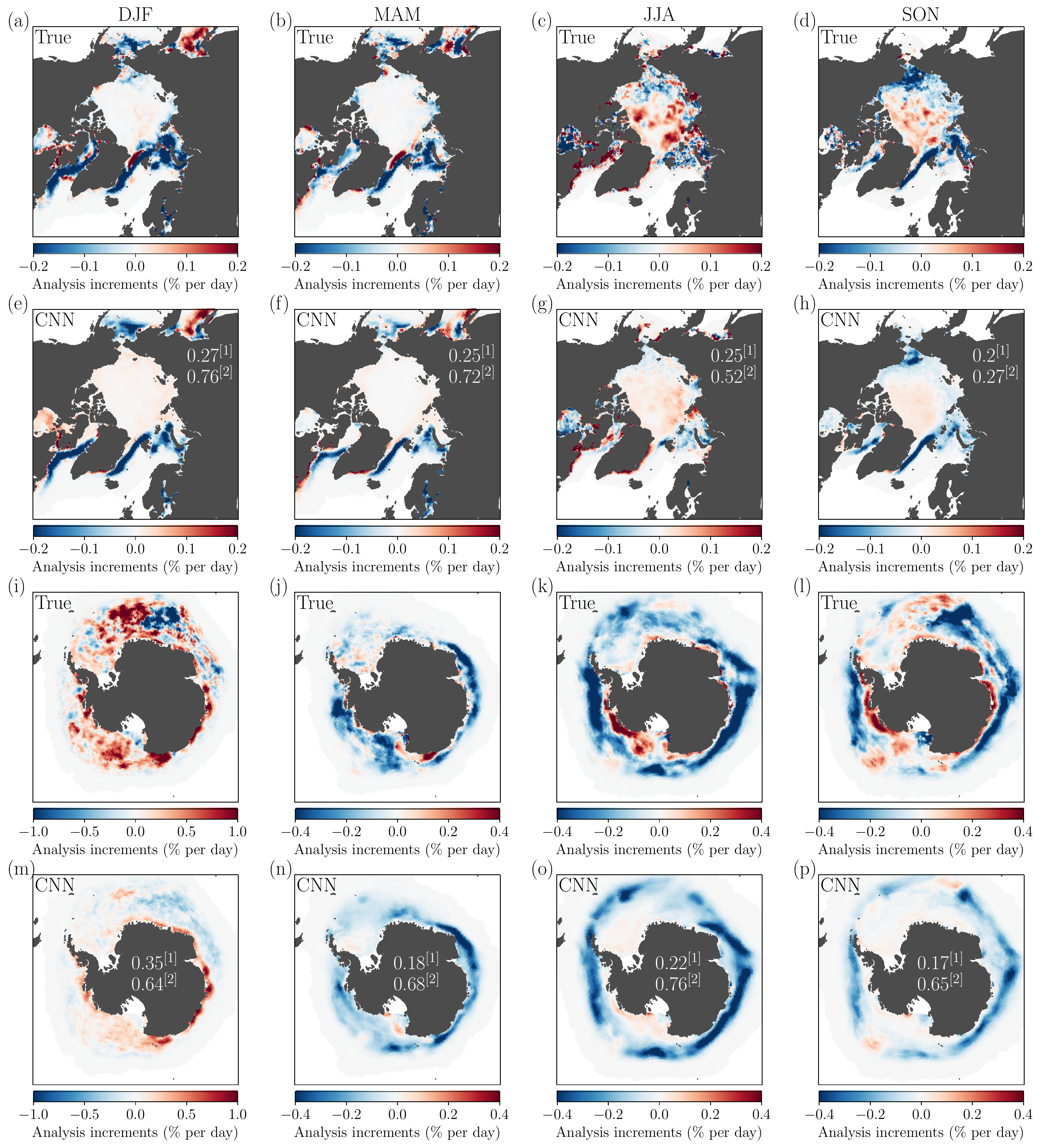}
    \caption{Seasonal climatologies of the (true) SPEAR category 3 sea ice concentration analysis increments and the equivalent CNN predictions, for both the Arctic (a)--(h) and Antarctic (i)--(p). Columns from left to right show DJF, MAM, JJA, SON climatologies, computed over the period 1982--2017. Values with the superscript [1] are the average of daily spatial pattern correlations between $\Delta$SICN\textsuperscript{True} and $\Delta$SICN\textsuperscript{CNN} in each respective season, while values with [2] are the spatial pattern correlations between the respective climatologies of the true and predicted increments.}
\label{fig:S6}
\end{figure}

\begin{figure}[t!]
    \centering
    \includegraphics[width=1\linewidth]{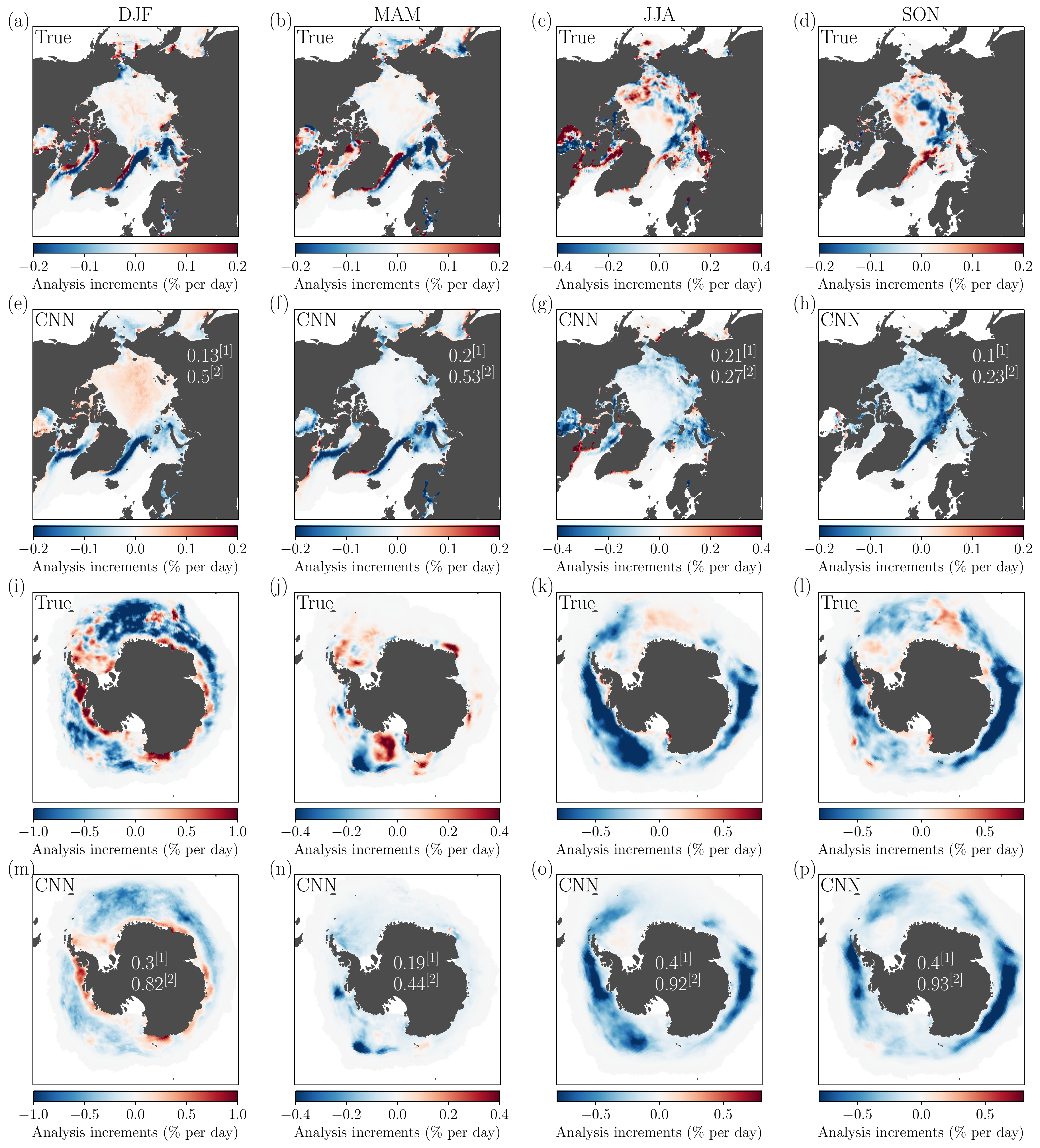}
    \caption{Seasonal climatologies of the (true) SPEAR category 4 sea ice concentration analysis increments and the equivalent CNN predictions, for both the Arctic (a)--(h) and Antarctic (i)--(p). Columns from left to right show DJF, MAM, JJA, SON climatologies, computed over the period 1982--2017. Values with the superscript [1] are the average of daily spatial pattern correlations between $\Delta$SICN\textsuperscript{True} and $\Delta$SICN\textsuperscript{CNN} in each respective season, while values with [2] are the spatial pattern correlations between the respective climatologies of the true and predicted increments.}
\label{fig:S7}
\end{figure}

\begin{figure}[t!]
    \centering
    \includegraphics[width=1\linewidth]{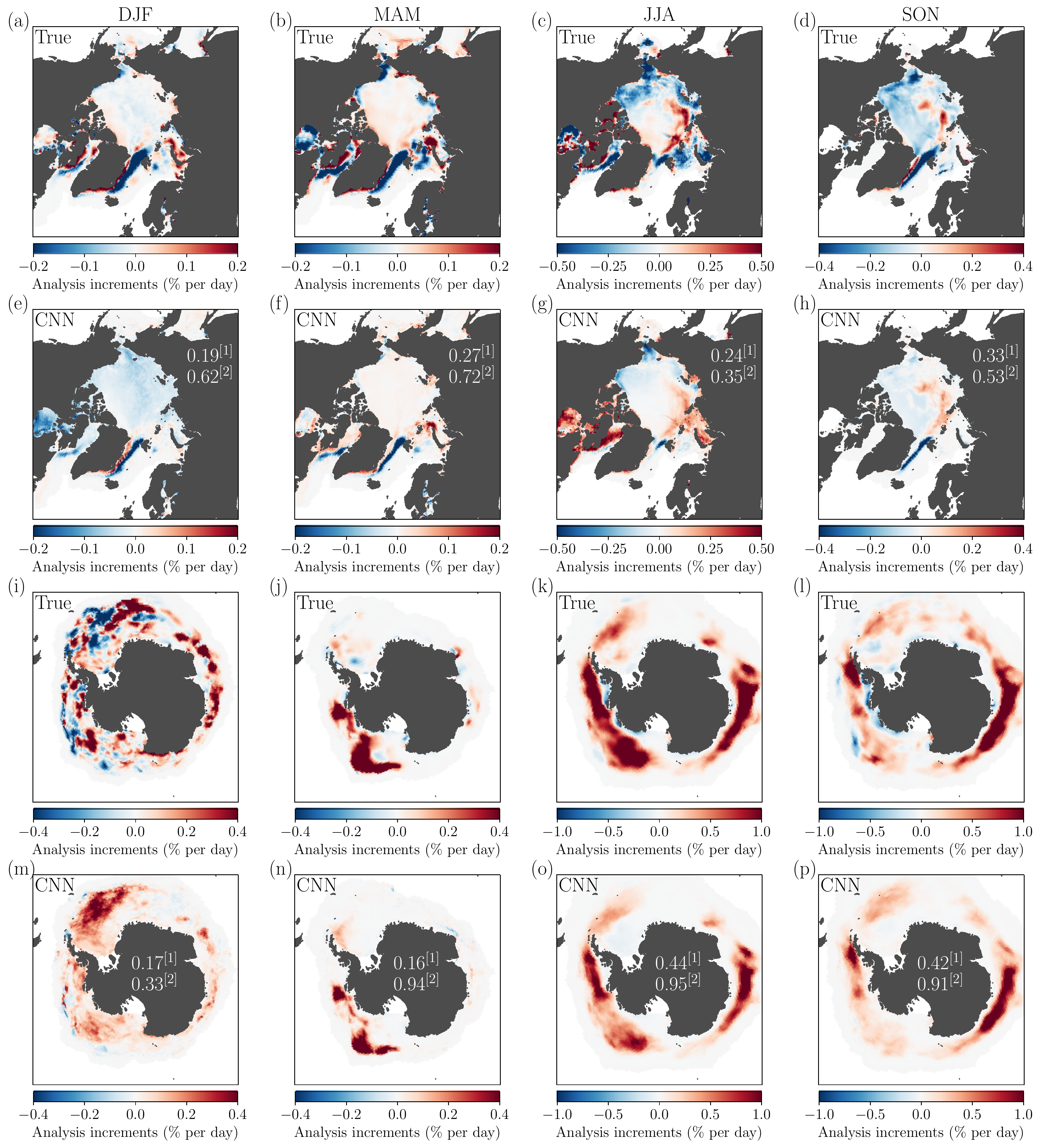}
    \caption{Seasonal climatologies of the (true) SPEAR category 5 sea ice concentration analysis increments and the equivalent CNN predictions, for both the Arctic (a)--(h) and Antarctic (i)--(p). Columns from left to right show DJF, MAM, JJA, SON climatologies, computed over the period 1982--2017. Values with the superscript [1] are the average of daily spatial pattern correlations between $\Delta$SICN\textsuperscript{True} and $\Delta$SICN\textsuperscript{CNN} in each respective season, while values with [2] are the spatial pattern correlations between the respective climatologies of the true and predicted increments.}
\label{fig:S8}
\end{figure}

\begin{figure}[t!]
    \centering
    \includegraphics[width=1\linewidth]{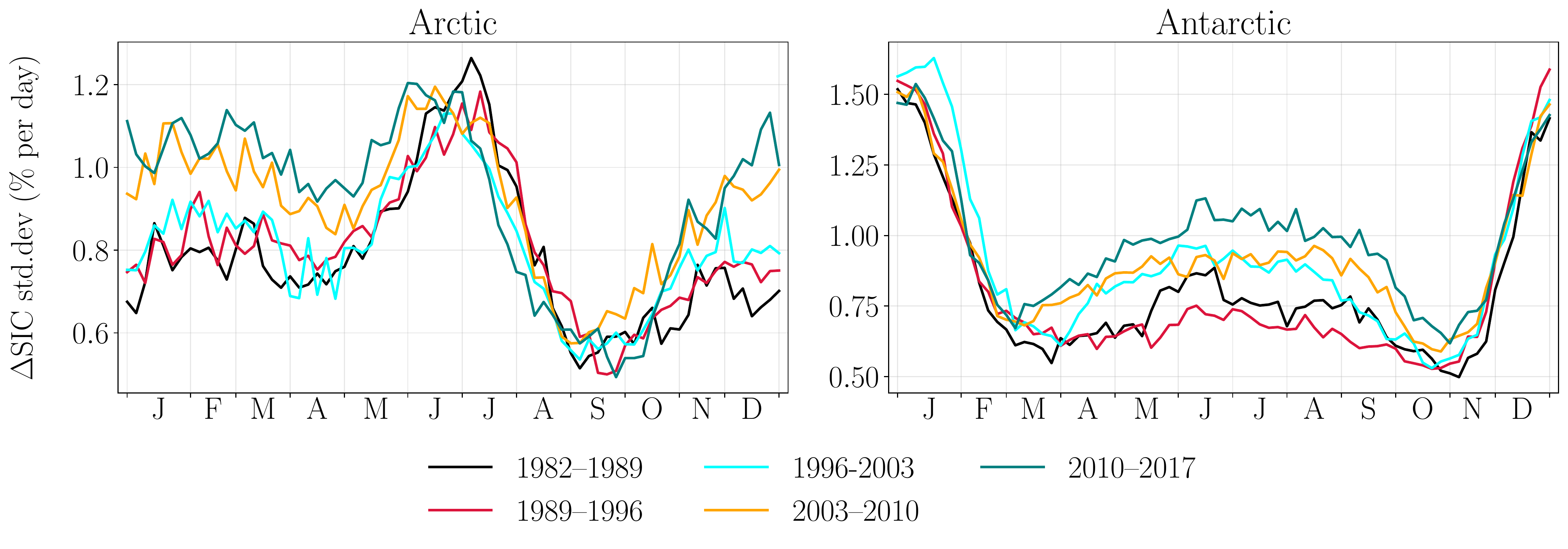}
    \caption{The standard deviation of the true aggregate sea ice concentration analysis increments ($\Delta$SIC\textsuperscript{True}), computed over each of the 5 cross-validation periods used for validating the CNN predictions. Shown as daily climatologies.}
\label{fig:S9}
\end{figure}

\begin{figure}[t!]
    \centering
    \includegraphics[width=1\linewidth]{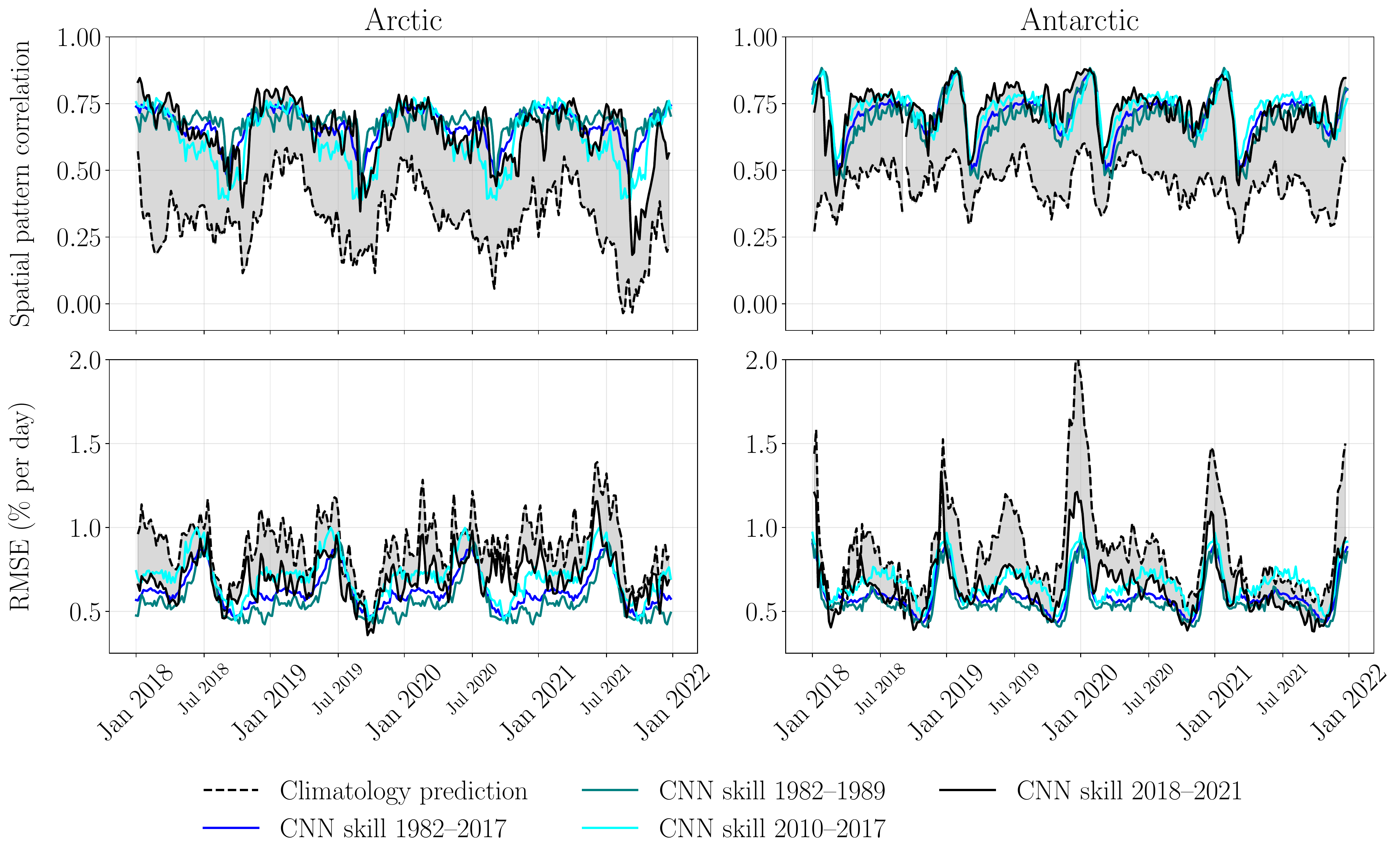}
    \caption{As in Figure 8 from the main article, except now highlighting the effect of non-stationarity within the increments by also including the climatological prediction skill for cross-validation chunks corresponding to the beginning of the time series record (1982--1989), as well as the end of the time series record (2010--2017).}
\label{fig:S10}
\end{figure}

\end{document}